\documentclass[12pt,titlepage]{article}
\usepackage[english]{babel}
\usepackage[dvips]{graphicx}

\begin{document}
\begin{titlepage}
\vspace{.5cm}

\begin{center}
{\Large \bf Cabibbo allowed $D \to  K \pi  \gamma$ decays\\}

\vspace{.5cm}

{\large \bf S. Fajfer$^{a,b}$, A. Prapotnik$^{b}$ and
P. Singer$^{c}$\\}

\vspace{0.5cm}
{\it a) Department of Physics, University of Ljubljana,
Jadranska 19, 1000 Ljubljana, Slovenia}

{\it b) J. Stefan Institute, Jamova 39, P. O. Box 300, 1001 Ljubljana,
Slovenia}\vspace{.5cm}

{\it c) Department of Physics, Technion - Israel Institute  of Technology,
Haifa 32000, Israel\\}

\end{center}

\vspace{1.5cm}

\centerline{\large \bf ABSTRACT}

\vspace{0.5cm}

The weak radiative Cabibbo allowed decays
$D^+ \to {\bar K}^0 \pi^+ \gamma$ and
$D^0 \to K^- \pi^+ \gamma$ with nonresonant $K \pi$ are 
investigated by relying
 on the factorization approximation for the nonleptonic
weak transitions  and  the model
which combines  the heavy quark effective theory and the
chiral Lagrangian approach.
The dominant contributions to the amplitudes come from the
long distance effects.
The decay amplitude has
both parity violating and parity conserving parts.
The parity violating part includes also a bremsstrahlung 
contribution.
The  branching ratio obtained for  the parity conserving part is of the order
$10^{-4}$ for the $D^0 \to K^- \pi^+ \gamma$ decay and 
$10^{-5}$ for $D^+ \to {\bar K}^0 \pi^+ \gamma$, when the 
effect of light vector mesons is included, and 
smaller otherwise.
 The branching ratio for 
the parity violating part with  a photon energy 
cut of $50$ ${\rm MeV}$, 
is close to 
$10^{-3}$ for the $D^0$ decay and $4 \times10^{-4}$ for the $D^+$ decay. 
We present Dalitz plots and energy spectra 
for both transitions as derived from our model and we 
probe the role of the light vector mesons in these decays.

\end{titlepage}

\setlength {\baselineskip}{0.55truecm}
\setcounter{footnote}{1}    

\setcounter{footnote}{0}
{\bf I. INTRODUCTION}\\

The investigation of radiative and dilepton weak decays of pseudoscalar 
charm mesons has been pursued rather vigorously in recent 
years, both theoretically and experimentally. To a certain extent, this 
activity has been fueled by the ongoing search for 
physics beyond the standard model, which might be of measurable 
consequence in certain charm radiative and dilepton decays 
[1-4].
To date, no radiative or dilepton weak decay of $D$ has been detected.
However, upper bounds have been established for a sizable number 
of these decays. The radiative decays $D^0 \to \rho^0, $ 
$\omega^0,$ $ \phi, $ $ \bar K^{*0}$ $+ \gamma$ were recently 
bounded \cite{CLEO1} to branching ratios in the $10^{-4}$ range,
which is approaching the standard model expectations (see, e.g.
\cite{FPS2,L} where additional previous works are mentioned). 
The dilepton decays $D \to P l^+ l^-$, $D \to V  l^+ l^-$ are the
subject of intensive searches at CLEO and Fermilab \cite{CLEO2}. 
Here again, with upper bounds of $10^{-5} - 10^{-4}$ 
for branching ratios of the various modes one approaches the 
expectations of the standard model 
\cite{BGHP0,FPS1,Sasathesis,FPS3}. The situation should improve in 
the 
future, due to new possibilities for observation of charm meson 
decays at 
BELLE, BABAR and Tevatron. Recently, upper limits in the 
$10^{-5} - 10^{-4}$ range were established \cite{E791} 
also for $D^0$ dilepton decays with two nonresonant pseudoscalar 
mesons in the 
final state $D^0 \to(\pi^+ \pi^-, K^- \pi^+, K^+ K^-)$ 
$ \mu^+ \mu^-$, though no comparable results are available yet for similar 
photonic decays.

In the present work, we undertake the study 
of the Cabibbo allowed radiative decays 
$D^+ \to {\bar K}^0 \pi^+ \gamma$ and
$D^0 \to K^- \pi^+ \gamma$, which we consider to be the most  
likely candidates for early detection. 
These decays are the charm sector counterpart of 
the $K \to \pi \pi \gamma$ \cite{KPPG1,KPPG2,KPPG3} decays, 
which have 
provided a wealth of information on meson dynamics. In the strange 
sector, the  $K^+ \to \pi^+ \pi^0 \gamma$ and 
 $K_L \to \pi^+ \pi^- \gamma$ are singled out as the most 
 suitable ones for the investigation of the 
 radiative decay mechanism; this, 
 since the relative suppresion of the corresponding  
 $K^+ \to \pi^+ \pi^0$ and $K_L \to \pi^+ \pi^-$ amplitudes leads to a
situation where the direct radiative transition is not
overwhelmed by the bremsstrahlung 
part. 

In the $K \to \pi \pi \gamma$ decays, the long - distance 
contribution is dominant \cite{KPPG3}. 
In the charm radiative decays, the theoretical studies show 
that likewise, the long - distance is the dominant feature of the decays 
\cite{BGHP0,Sasathesis,FPS2,L,FPS3}. The short - distance 
contribution realized by the penguin diagram $c \to u \gamma$ 
\cite{x.y.pham,GHMW,BGHP} might play a role in certain Cabibbo 
suppressed decays, which are not discussed in the present paper. 

In the charm sector, the nonleptonic D meson decays still 
provide a continuing theoretical challenge 
(see, e.g. \cite{BFOS,KSUV} and references therein).
The short distance effects are considered well understood 
but the perturbative techniques required for the
evaluation of certain matrix elements are based on approximate models. 
Usually the factorization approximation is used (see, e.g.  
\cite{BFOS,factor}), although 
the experimental data indicate the apparent need for the inclusion 
of nonfactorizable amplitudes in certain channels. 

In this first treatment of $D \to K \pi \gamma$ 
decays we  use the factorization approximation for the 
calculation of weak transition elements. We consider 
the use of this  approach to be justified by the "near" success of the approach for
the nonleptonic amplitudes. This will involve its use in the $D(D^*) K\pi$  vertices 
as well as in $D(D^*) \to V$ and $D(D^*) \to P$ transitions, 
 all of which required for the calculations of the 
 $D \to K \pi \gamma$ amplitudes within our model. 
 For the evaluation of the $(D,D^*) \to (P,V)$ transitions, we use the 
 information obtained for these matrix elements from semileptonic 
 decays (see, e.g. \cite{BFO2}). The general theoretical 
 framework for our calculation is that of the heavy quark chiral 
 Lagrangian \cite{wise,YAN}. 
 In the $K \to \pi \pi \gamma$ decays, it has been shown that 
 intermediate light vector mesons play an important role in the decay 
 amplitude \cite{KPPG1}. We shall investigate the role of 
 intermediate light vector mesons also in the $D \to K \pi \gamma$ amplitude. 
In order to accomplish this, we use the extension of the formalism 
of \cite{wise,YAN} to include also the light vector mesons 
\cite{castwo,BFO1}. 

The present study of $D^+ \to {\bar K}^0 \pi^+ \gamma$ and
$D^0 \to K^- \pi^+ \gamma$ shows that the direct part of the 
radiative amplitude is not much smaller in strength than the bremsstrahlung 
part, rather similarly to the inhibited K decays mentioned above. 
If confirmed by experiments, it places these decays in the status 
of a most suitable ground for the investigation 
of the mechanisms involved in such nonleptonic D decays.

In Section II we present the theoretical framework for our calculation.
In Section III we display the explicit expressions of all the 
calculated decay amplitudes.  Section IV 
contains the discussion and the summary. \\

\vskip 1cm
{\bf II. THE THEORETICAL FRAMEWORK}\\

The nonradiative two-body D - decays, from which the bremsstrahlung part
of the radiative decays originates, are 
$D^0 \to K^- \pi^+$  and $D^+ \to \bar K^0 \pi^+$. The weak 
$\Delta I = 1$ transition leads to two independent isospin 
amplitudes in the final state, $A_{1/2}$ and $A_{3/2}$
and the relations to the physical decays is \cite{KSUV} 
\begin{equation}
A(D^+ \to \bar K^0 \pi^+) = A_{3/2}; \enspace A(D^0 \to K^- \pi^+)
=\frac{2}{3}A_{1/2} + \frac{1}{3}A_{3/2}.
\label{a}
\end{equation} 
From the determined branching ratios \cite{PDG} of 
$BR(D^+ \to \bar K^0 \pi^+) =(2.89\pm 0.26)\%$ and 
$BR(D^0 \to K^- \pi^+)=(3.83\pm 0.09)\%$ 
one learns that the relative size of the absolute values of the 
amplitudes $|A(D^0 \to K^- \pi^+)|/|A(D^+ \to \bar K^0 \pi^+)|$ is 
$1.84$. Using also the information
from the third decay, 
$A(D^0 \to \bar K^0 \pi^0)$ $= \frac{{\sqrt 2}}{3}A_{1/2} - 
\frac{{\sqrt 2}}{3}A_{3/2}$, the isospin analysis shows 
that  $|A_{1/2}|/|A_{3/2}|\simeq 2.7$, and their relative phase is 
$90^o$ \cite{Rosner}.  Despite this knowledge, there is still no 
complete interpretation for the mechanisms leading to the decays 
\cite{Rosner}, although it is clear that the situation 
is different from the $ K \to \pi \pi $ channels, where 
$\Delta I =1/2$ enhancement introduces a large disparity between the 
final state isospin amplitudes. The relevance of the above 
picture to the radiative decays 
will be discussed in the last Section.

Since our problem of describing the $ D \to K \pi \gamma$ 
decays involves transitions between heavy mesons 
and light pseudoscalars, we adopt the effective Lagrangian 
\cite{wise,YAN} which contains both the 
heavy flavor and the $SU(3)_L \times SU(3)_R$ 
chiral symmetry as the theoretical framework for our calculation. 
From the experience with $K \to \pi \pi \gamma$ decays, one knows  
\cite{KPPG1,KPPG2,KPPG3} that the decay amplitude is largely determined by 
contributions from virtual vector mesons. Considering the possibilty 
that vector mesons would play a role in the $D \to K \pi \gamma$ 
decays as well 
(we remind the reader that we consider here nonresonant $K \pi$, 
the decays $D \to K^* \gamma$ having been treated 
separately \cite{BGHP,FS}), we should complement the 
Lagrangian by introducing light vector mesons. For this 
we choose the generalization of the original Lagrangian 
\cite{wise,YAN} by Casalbuoni et al \cite{castwo} in which the 
original symmetry is broken spontaneously to diagonal  $SU(3)_V$ \cite{bando} with the introduction of the 
light vector mesons. We present here in some detail this 
formalism (for more details see \cite{FS}), which we use as the main 
tool of our calculation. We shall perform the 
calculations also without vector mesons 
in the Lagrangian in which case the original heavy quark chiral 
Lagrangian \cite{wise,YAN} is used, in order to clarify their  role in 
these decays. 

The light degrees of freedom are described by the
3$\times$3 Hermitian matrices

\begin{eqnarray}
\label{defpi}
\Pi = \pmatrix{
{\pi^0 \over \sqrt{2}} + {\eta_8 \over \sqrt{6}}+ {\eta_0 \over \sqrt{3}}&
\pi^+ & K^+ \cr
\pi^- & {-\pi^0 \over \sqrt{2}} + {\eta_8 \over \sqrt{6}} +
{\eta_0 \over \sqrt{3}} & K^0 \cr
K^- & {\bar K^0} & -{2 \eta_8 \over \sqrt{6}} + {\eta_0 \over \sqrt{3}} \cr}
\end{eqnarray}

\noindent
and

\begin{eqnarray}
\label{defrho}
\rho_\mu = \pmatrix{
{\rho^0_\mu + \omega_\mu \over \sqrt{2}} & \rho^+_\mu & K^{*+}_\mu \cr
\rho^-_\mu & {-\rho^0_\mu + \omega_\mu \over \sqrt{2}} & K^{*0}_\mu \cr
K^{*-}_\mu & {\bar K^{*0}}_\mu & \Phi_\mu \cr}
\end{eqnarray}

\noindent
for the pseudoscalar and vector mesons, respectively. They are
usually expressed through the combinations

\begin{eqnarray}
\label{defu}
u & = & \exp  \left( \frac{i \Pi}{f} \right)\;,
\end{eqnarray}

\noindent
where $f\simeq f_{\pi}=132$ MeV is the pion pseudoscalar decay constant, and

\begin{eqnarray}
\label{defrhohat}
{\hat \rho}_\mu & = & i {g_v \over \sqrt{2}} \rho_\mu\;,
\end{eqnarray}

\noindent
where $g_v=5.9$ was fixed in the case of exact flavor symmetry
\cite{bando}.
In the following we will also use the gauge field tensor $F_{\mu \nu}({\hat \rho})$

\begin{eqnarray}
F_{\mu \nu} ({\hat \rho}) =
\partial_\mu {\hat \rho}_\nu -
\partial_\nu {\hat \rho}_\mu +
[{\hat \rho}_\mu,{\hat \rho}_\nu].
\label{deff}
\end{eqnarray}
It is convenient to introduce two currents
${\cal V}_{\mu} =  \frac{1}{2} (u^{\dag}
D_{\mu} u + u D_{\mu}u^{\dag})$
and ${\cal A}_{\mu}  =  \frac{1}{2} (u^{\dag}D_{\mu} u - u D_{\mu}u^{\dag})$.
The covariant derivative of $u $ and $u^{\dag}$ is
defined as $D_{\mu}u = (\partial_{\mu} + {\hat B}_{\mu} )u$
and $D_{\mu}u^{\dag}= (\partial_{\mu} + {\hat B}_{\mu} )u^{\dag}$,
with ${\hat B}_{\mu} = i e B_{\mu} Q$,
$Q = diag (2/3,-1/3,-1/3)$, $B_{\mu}$ being the photon field.

The light meson part of the strong
Lagrangian can be written as \cite{bando}
\begin{eqnarray}
\label{deflight}
{\cal L}_{light} & = & -{f^2 \over 2}
\{tr({\cal A}_\mu {\cal A}^\mu) +
a\, tr[({\cal V}_\mu - {\hat \rho}_\mu)^2]\}\nonumber\\
& + & {1 \over 2  g_v^2} tr[F_{\mu \nu}({\hat \rho})
F^{\mu \nu}({\hat \rho})]\;.
\end{eqnarray}
The constant $a$ in (\ref{deflight}) is in principle a free parameter.
In the case of exact vector meson dominance (VDM) $a=2$ \cite{bando,fuj}.
However, the photo production and decays data indicate \cite{EIMS}
that the $SU(3)$ breaking modifies the VDM in
\begin{eqnarray}
\label{VDM}
{\cal L}_{V-\gamma} = -e g_v f^2 B_{\mu} (\rho^{0\mu} + \frac{1}{3}
\omega^{\mu} - \frac{{\sqrt 2}}{3} \Phi^{\mu}).
\end{eqnarray}
Instead of the exact $SU(3)$ limit ($g_v= m_V/f$),
we shall use  the measured values, defining
\begin{eqnarray}
< V(\epsilon_V,q)| V_{\mu} |0> = i \epsilon_{\mu}^{*} (q) g_V(q^2).
\label{defg}
\end{eqnarray}
The couplings $g_V(m_V^2)$ are obtained from the leptonic decays of these
mesons.
In our calculation we use
$g_{\rho}(m_{\rho}^2) \simeq g_{\rho}(0)= 0.17 GeV^2$,
$g_{\omega}(m^2_{\omega}) \simeq g_{\omega} (0) = 0.15 GeV^2$
and $g_{\Phi}(m^2_{\Phi}) \simeq g_{\Phi} (0) = 0.24 GeV^2$.

Both the heavy pseudoscalar and the heavy vector
mesons are incorporated in a $4\times 4$ matrix
\begin{eqnarray}
\label{defh}
H_a& = & \frac{1}{2} (1 + \!\!\not{\! v}) (P_{a\mu}^{*}
\gamma^{\mu} - P_{a} \gamma_{5})\;,
\end{eqnarray}
\noindent
where $a=1,2,3$ is the $SU(3)_V$ index of the light
flavors, and $P_{a\mu}^*$, $P_{a}$, annihilate a
spin $1$ and spin $0$ heavy meson $Q \bar{q}_a$ of
velocity $v$, respectively. They have a mass dimension
$3/2$ instead of the usual $1$, so that the Lagrangian
is in the heavy quark limit $m_Q\to\infty$ explicitly
mass independent. Defining moreover
\begin{eqnarray}
\label{defhbar}
{\bar H}_{a} & = & \gamma^{0} H_{a}^{\dag} \gamma^{0} =
(P_{a\mu}^{* \dag} \gamma^{\mu} + P_{a}^{\dag} \gamma_{5})
\frac{1}{2} (1 + \!\!\not{\! v})\;,
\end{eqnarray}

\noindent
we can write the strong  Lagrangian as \cite{BFO1}
\begin{eqnarray}
\label{deflstrong}
{\cal L}_{even} & = & {\cal L}_{light} +
i Tr (H_{a} v_{\mu} D^{\mu}
{\bar H}_{a})\nonumber\\
& + &i g Tr [H_{b} \gamma_{\mu} \gamma_{5}
({\cal A}^{\mu})_{ba} {\bar H}_{a}]
 +  i {\tilde \beta }Tr [H_{b} v_{\mu} ({\cal V}^{\mu}
- {\hat \rho}_{\mu})_{ba} {\bar H}_{a}]\;,
\end{eqnarray}
where
$ D^{\mu} {\bar H}_{a} = (\partial_{\mu} + {\cal V}_{\mu} -
i e Q^{\prime} B_{\mu}){\bar H}_{a}$, with $Q^{\prime} = 2/3$
for c  quark.

The coupling $g$ can be fixed \cite{GS} by using the data  
\cite{CLEO} on 
$D^* \to D \pi$ decay width. These data give $g =0.59$.
The plus sign is taken to be in agreement with the quark model studies.
The parameter  ${\tilde \beta}$ is less known, but it seems that it
can be safely neglected \cite{BFOS}.

The electromagnetic field can couple to the mesons also through the anomalous
interaction; i.e., through the odd parity Lagrangian.
The contributions to this Lagrangian arise from terms  of the Wess - Zumino -
Witten
kind,  given by \cite{fuj,BGP}

\begin{eqnarray}
\label{defloddlight}
{\cal L}^{(1)}_{odd} & = & -4 \frac{C_{VV\Pi}}{f} \epsilon
^{\mu \nu \alpha \beta}Tr (\partial_{\mu}
{\rho}_{\nu} \partial_{\alpha}{\rho}_{\beta} \Pi).
\end{eqnarray}

The coupling $C_{VV\Pi}$ can be
determined in the case of the exact $SU(3)$ flavor symmetry
following the hidden symmetry approach of \cite{bando,fuj} and
it is found to be $C_{VV\Pi} = 3g_v^2 /32 \pi^2 = 0.33$.
In the actual calculation, we allowed for $SU(3)$ symmetry 
breaking and we used the $VP\gamma$ coupling as determined from 
experiment \cite{PDG}. 
We will also need the odd - parity  Lagrangian in the heavy sector.
Such terms are required by the $D^* \to D \gamma$ transition, which 
cannot be generated from (\ref{deflstrong}).  
There are two contributions \cite{BFO1,CFN} in it,
characterized by coupling strengths $\lambda$ and $\lambda^{\prime}$.
The first is given by
\begin{eqnarray}
\label{defoddheavy}
{\cal L}_{1} & = & i {\lambda} Tr [H_{a}\sigma_{\mu \nu}
F^{\mu \nu} (\hat \rho)_{ab} {\bar H_{b}}].
\end{eqnarray}
In this term the interactions of light vector mesons with heavy pseudoscalar
or heavy vector mesons is described. The light vector meson
can then couple to the
photon by the standard VDM prescription.
This term is of the order ${1}/{{\lambda}_{\chi}}$ with
${\lambda}_{\chi}$ being the chiral perturbation theory scale 
\cite{CG}.

The second term gives the direct heavy quark - photon interaction and
is generated by the Lagrangian
\begin{eqnarray}
\label{defoddheavy2}
{\cal L}_{2} & = & - {\lambda}^{\prime} Tr [H_{a}\sigma_{\mu \nu}
F^{\mu \nu} (B) {\bar H_{a}}].
\end{eqnarray}
The parameter ${\lambda}^{\prime} $ is  
given in heavy quark symmetry 
limit by 
${\lambda}^{\prime} \simeq - 1/(6 m_c)$
\cite{YAN} and it should be considered as a higher order
term in $1/m_Q$ expansion \cite{S}.

In order to gain information on these couplings one has to use the
existing data on
$D^{*0} \to D^0 \gamma$,
$D^{*+}\to D^+ \gamma$ and $D_s^{*+} \to D^+_s \gamma$ decays.
Experimentally, the ratios
$R_{\gamma}^0 = \Gamma (D^{*0} \to D^0 \gamma)/\Gamma
(D^{*0} \to D^0 \pi^0)$ and
$R_{\gamma}^+ = \Gamma (D^{*+} \to D^+ \gamma)/\Gamma
(D^{*+} \to D^+ \pi^0)$ are known \cite{PDG}.
These data determine two possibilities \cite{FS}.
One of them is $|\lambda /g| = 0.839$  $GeV^{-1}$,
$|\lambda^{\prime} /g |= 0.175$  $GeV^{-1}$.  The second one does not
agree with present data.
With $g=0.59$ we obtain
$\lambda =\pm 0.49$ $GeV^{-1}$ 
and $\lambda' = \pm0.102$ $GeV^{-1}$.

The  $\lambda'\simeq - 1/(6 m_c)$ would give with the mass of 
charm quark  $m_c = 1.4$  
$ GeV$ that $\lambda' = -0.12$ $GeV^{-1}$,   in good agreement with 
the above value. The simple quark model analysis 
indicates that  $\lambda'$ and $\lambda$ 
are both negative \cite{S}. In our numerical calculations 
we give the results using these parameters. 
In the literature (e.g. \cite{GS,S,FSZ}) instead of $\lambda$
the $\beta$ parameter is often used. 
The value $\beta = 2.3$ $GeV^{-1}$ corresponds to $\lambda=-0.49$ 
$GeV^{-1}$, since $2 \lambda \frac{g_v}{{\sqrt 2}}$ 
$(g_{\rho}/m_{\rho}^2 + g_{\omega}/(3m_{\omega}^2)) = - (2/3) \beta$.

In addition to strong and electromagnetic interactions, we have to
specify the weak one. The nonleptonic weak Lagrangian on the
quark level for the Cabibbo allowed decays can be written as usual
\cite{factor}
\begin{eqnarray}
\label{deflfermisl}
{\cal L}_{NL}^{eff}(\Delta c=\Delta s=1)
& = & -{G_F \over \sqrt{2}} V_{ud}V_{cs}^*
[ a_1 O_1 +  a_2 O_2 ],
\end{eqnarray}
\noindent
where $O_1 = ({\bar u} d)^{\mu}_{V-A}
({\bar s}c )_{V-A,\mu}$ and $O_2 = ({\bar u} c)_{V-A,\mu}
({\bar s} d)_{V-A}^{\mu}$,  
$V_{ij}$ are the CKM matrix elements, $G_F$ is the
Fermi constant and $({\bar \Psi}_1\Psi_2)^\mu\equiv
{\bar \Psi}_1\gamma^\mu(1-\gamma^5)\Psi_2$.
In our calculation we use $a_1 = 1.26$ and $a_2 = -0.55$ as
found in \cite{factor}.

At the hadronic level, the weak current transforms as $({\bar 3}_L,1_R)$
under chiral $SU(3)_L\times SU(3)_R$, is linear in the
heavy meson fields $D^a$ and $D^{*a}_\mu$ and is taken as  \cite{BFO2}

\begin{eqnarray}
\label{jqbig}
{J_Q}_{a}^{\mu} = &\frac{1}{2}& i \alpha Tr [\gamma^{\mu}
(1 - \gamma_{5})H_{b}u_{ba}^{\dag}]\nonumber\\
&+& \alpha_{1}  Tr [\gamma_{5} H_{b} ({\hat \rho}^{\mu}
- {\cal V}^{\mu})_{bc} u_{ca}^{\dag}]\nonumber\\
&+&\alpha_{2} Tr[\gamma^{\mu}\gamma_{5} H_{b} v_{\alpha}
({\hat \rho}^{\alpha}-{\cal V}^{\alpha})_{bc}u_{ca}^{\dag}]+...\;,
\end{eqnarray}

\noindent
where $\alpha=f_H\sqrt{m_H}$ \cite{wise}, $\alpha_1$ was
first introduced by Casalbuoni et al. \cite{castwo}, while $\alpha_2$
was introduced in \cite{BFO2}.
It has to be included,
since it is of the same order in the $1/m_Q$ and chiral
expansion as the term proportional to $\alpha_1$ \cite{BFO2}. \\
The relevant
matrix element is parametrized usually in
$D \to V l \nu_l$ semileptonic decay as \cite{BGHP,factor,BFO2,RB}

\begin{eqnarray}
\label{parhv}
<V(p_V,\epsilon_V)|(V-A)^\mu|D(p)>=
{2 V(q^2)\over m_D+m_V}
\epsilon^{\mu\nu\alpha\beta}\epsilon_{V\nu}^* p_\alpha
{p_V}_\beta \nonumber\\
+i \epsilon^*_V.q {2 m_V\over q^2}q_\mu ( A_3(q^2) - A_0(q^2))
+i(m_D+m_V)[\epsilon_{V\mu}^* A_1(q^2) \nonumber\\
-{\epsilon^*_V .q\over m_D+m_V}((p+p_V)_\mu
A_2(q^2)] \;,
\end{eqnarray}
where $q = p - p_V$. In order that these matrix elements should
be finite at $q^2 = 0$, the form factors satisfy the relation
\cite{factor}

\begin{equation}
\label{relff}
A_3(q^2)-{m_H+m_V\over 2 m_V}A_1(q^2)+
{m_H-m_V\over 2 m_V}A_2(q^2)=0\;,
\end{equation}
and $A_3 (0) = A_0(0)$.
We take the  following expressions for the
form factors at $q^2_{max}$ \cite{castwo} 
\begin{equation}
V(q_{max}^2)=\frac{1}{\sqrt{2}}\lambda g_v f_D
\frac{M+m}{M+\Delta},
\label{Vq}
\end{equation}
\begin{equation}
A_1(q_{max}^2)=\frac{-\sqrt{2}\alpha_1g_v\sqrt{M}}{M+m},
\label{A1q}
\end{equation}
and
\begin{equation}
A_2(q_{max}^2)=-\frac{2g_v}{\sqrt{2}} 
\frac{M+m}{M^{3/2}}\alpha_2,
\label{A2q}
\end{equation}
where $\Delta$ stands for the $D^*$ and $D$ mass difference. 
Assuming the pole dominance one can connect the value of form factors
at $q^2_{max}$ and $0$ momentum transfer by 
$ F(0)=F(q_{max})\left(1-{(M-m)^2}/{M_p^2}\right)$, 
where $F$ stands for $V$, $A_1$ or $A_2$,   $M$ is the  
D meson mass and $m$ is the light vector meson mass. 
Using the experimental data \cite{PDG}
$|V^{DK^*} (0)| = 1.02 \pm 0.12$, $|A_1^{DK^*} (0)| = 0.55 \pm0.03$ and
$|A_2^{DK^*}(0)| = 0.40 \pm 0.07$,  we find for the  couplings
$\lambda=-0.56  $ ${\rm GeV}^{-1}$,
$|\alpha_1|=0.156$ ${\rm GeV}^{1/2}$, $|\alpha_2|=0.052
$${\rm GeV}^{1/2}$. 
The value of $\lambda$ is in  good agreement with results obtained from
$D^* \to D \gamma $ data.

The light weak current is derived to be \cite{BFO1}
\begin{equation}
J^{\mu}_{ij} = i f^2 \{u [ {\cal A}^\mu - a ({\cal V}^\mu -
{\hat \rho}^\mu] u^\dag\}_{ji}.
\label{lwc}
\end{equation}

The photon emission is obtained by gauging the weak sector too.
The important consequence of this procedure is that 
thereby the gauge
invariance of the
whole amplitude is achieved. 
This turns out to be equivalent to the usual procedure of achieving 
gauge invariance in bremsstrahlung processes with a momentum 
dependent strong vertex, as pointed out \cite{Singer}
for the somewhat similar process $V \to PP' \gamma$. 
Actually by gauging the weak
sector we produce the same
graphs, which were necessary to induce to satisfy the gauge
invariance \cite{Singer}.  \\

{\bf III. THE DECAY AMPLITUDES}\\

The general Lorentz decomposition of the $D(P) \to K(p) \pi(q) \gamma
(k,\epsilon)$
decay amplitude is given by

\begin{equation}
{\cal M}=-\frac{G_f}{\sqrt{2}}V_{cs}V_{du}^*\left({\bar F}_1
\left[\frac{q \cdot\varepsilon}{q  \cdot k}-
\frac{p\cdot \varepsilon}{p\cdot k}\right]
+F_2 \varepsilon^{\mu \alpha \beta \gamma}
\varepsilon_\mu v_\alpha k_\beta q_\gamma \right).
\label{amp}
\end{equation}
The part of the amplitude containing the ${\bar F}_1$ 
form factor is parity violating,
while the 
one with $F_2$ is parity conserving. Both of them are 
functions of  scalar products of momenta as $k \cdot p$, 
$k\cdot q$. 
Note that ${\bar F}_1$ contains 
contributions  which arise from bremsstrahlung part of 
the amplitude
as well as  a direct electric transition. 
On the other hand, $F_2$ corresponds to the magnetic transition. 

In order to determine ${\bar F_1}$, $F_2$ we use the model described in
the previous Section. The diagrams contributing to these
form factors are given in Figures 1-4. In Figures 1 and 2 
 are given Feynman diagrams contributing  to 
 the  $D^+ \to \bar K^0 \pi^+ \gamma$ decay amplitude
while  the contributions to the $D^0 \to K^- \pi^+ \gamma$
decay amplitude  are presented in Figures 3 and 4.
Note that we denote  heavy mesons by one full line,
light pseudoscalar mesons by dashed lines,
light vector mesons by two full lines and photons by wavy lines.
The weak vertex is denoted by square box. 

Before proceeding to the actual calculation, we note the 
following complication. 
As well known, the leading terms of the expansion of the 
radiative amplitude in the  
photon momentum $(k)$ are determined \cite{Low} 
by the original amplitude ($ D \to K\pi$  in our case). 
However, the nonleptonic $ D \to K\pi$ amplitude cannot be calculated 
accurately in the factorization approximation from the diagrams 
provided by our model. 
Such a calculation gives a rather good result for the 
$D^+ \to {\bar K}^0 \pi^+$ channel but is less successful 
for the $D^0 \to K^+ \pi^-$ decay. 
In order to overcome this defficiency and to be able to present 
accurately the  bremsstrahlung component of the radiative transition, we 
shall use an alternative approach for its derivation. 
This approach then is to use the values of the experimental amplitudes 
$D \to K \pi$, assumed to have no internal structure, for the calculation of 
the bremsstrahlung component. In order to 
accomodate this we  
 rewrite the decay amplitude (24) as 
\begin{equation}
{\cal M} = -\frac{G_f}{\sqrt{2}}V_{cs}V_{du}^*\left(F_0
\left[\frac{q \cdot\varepsilon}{q  \cdot k}-
\frac{p\cdot \varepsilon}{p\cdot k}\right]
+ F_1 \left[ (q \cdot\varepsilon) (p  \cdot k)- 
(p \cdot\varepsilon) (q  \cdot k)\right]\nonumber\\
+ F_2 \varepsilon^{\mu \alpha \beta \gamma}
\varepsilon_\mu v_\alpha k_\beta q_\gamma \right),
\label{amp1}
\end{equation}
where $F_0$ is the experimentally determined $D \to K\pi$ 
amplitude and $F_1$, $F_2$ are the form factors of the 
electric and magnetic direct transitions which we calculate with our model.
When intermediate states appear to be on the mass shell,
we use Breit Wigner formula.
However, we remark at this point that since we are interested in 
the $D \to (K \pi)_{nonres} \gamma$ transitions, we delete the region of 
the $K^*$ resonance appearing in diagram $D_{1,2}^0$ and we retain 
only the region in $(p + q)^2$ which is beyond 
$m_{K^*} \pm \Gamma_{K^*}$.

In Appendix A 
we present explicitly expressions for the form factors 
for the decay $D^+ \to \bar K^0 \pi^+ \gamma$ using the notations
$A_i^+$, $B_i^+$, etc. .
The amplitude  $A_i^+$, $B_i^+$, etc. is obtained as a sum of the amplitudes
presented by the corresponding diagrams in the $i$-th row in 
Figures 1-4. Each $A_i^{+}$ (or $B_i^+$, etc) is gauge invariant.  
For the electric parity - violating transition, we define both 
the total amplitude provided by the model ${\bar F}_1$, as well
as the direct part only, $F_1$, obtained after deleting the bremsstrahlung 
diagrams. Then the amplitudes  for $D^+ \to \bar K^0 \pi^+ \gamma$
are 
\begin{equation}
{\bar F}_1(D^+ \to \bar K^0 \pi^+ \gamma)= \sum_{i=1}^{4}(
 A_i^{+}+ C_i^{+}),
\label{barF1+}
\end{equation}
\begin{equation}
F_1(D^+ \to \bar K^0 \pi^+ \gamma)= \frac{1}{(p\cdot k)(q \cdot k)}\sum_{i=3}^{4}(
 A_i^{+}+ C_i^{+}),
\label{F1+}
\end{equation}
\begin{equation}
F_2(D^+ \to \bar K^0 \pi^+ \gamma)= \sum_{i=1}^{3}( B_i^{+}+ 
 D_i^{+}).
\label{F2+}
\end{equation}
In the case of $D^0 \to K^- \pi^+ \gamma$ 
we have 
\begin{equation}
{\bar F}_1(D^0 \to K^- \pi^+ \gamma)= \sum_{i=1}^{4}( A_i^{0}+ 
C_i^{0}),
\label{barF10}
\end{equation}
\begin{equation}
F_1(D^0 \to K^- \pi^+ \gamma)=  
\frac{1}{(p\cdot k)(q \cdot k)}\sum_{i=3}^{4} A_i^{0}, 
\label{F10}
\end{equation} 
\begin{equation}
F_2(D^0 \to K^- \pi^+ \gamma)= \sum_{i=1}^{3}( B_i^{0}+  D_i^{0}),
\label{F20}
\end{equation}
where $A_i^0$, $B_i^0$ etc. are  gauge invariant sums of the 
amplitudes arising
from the graphs in the $i$-th 
row. 
In Appendix B we present the form factors for the
$D^0 \to K^- \pi^+ \gamma$ decay. 
We denoted by $A_i^{+,0}$, $B_i^{+,0}$  contributions which are created by $O_1$ 
operator and by $C_i^{+,0}$,  $D_i^{+,0}$ contributions caused by $O_2$. 

As we mentioned, in the calculation we used the experimental
value of $A(D\to K \pi)$ to calculate the bremsstrahlung part; $F_1$ is then
calculated by subtracting the bremsstrahlung component from 
$\bar F_1$. 

The differential cross section of the decays is given by 
\begin{equation}
d \Gamma = \frac{1}{(2 \pi)^3} \frac{1}{32 M^3} |{\cal M}|^2 
d m_{12}^2 d m_{23}^2, 
\label{gamma}
\end{equation}
where ${\cal M}$ is the decay amplitude, given by Eqs. (\ref{amp}) 
or  (\ref{amp1}), 
$m_{12}^2 = (P -k)^2$,
and $m_{23}^2 = (P- p)^2$, where $P$, $k$, $p$ are respective 
four - momenta of D- meson, photon and K meson. 
The total decay width is a sum of the parity conserving expressed by 
$F_2$ and the parity violating contributions expressed by 
$F_0$, $F_1$, $\Gamma = \Gamma_{PC} + \Gamma_{PV}$ 
(the PC and PV amplitudes do not interfere in the total width).
 Before giving numerical results, we make a few comments. 

The expressions for the amplitudes given in the Appendices 
contain several constants. A few are well determined 
(we use values given in \cite{PDG}) and 
require no further explanation; as to the rest, for $f_D$ we use the lattice result,
$f_D = 207$ ${\rm MeV}$ \cite{sahrajda} and for $f_{D^*}=1.13 f_D$.   
The couplings $g$, $\lambda$, $\lambda'$ are determined as previously 
explained and we use $g =0.59$, $\lambda = -0.49$ ${\rm GeV}^{-1}$
and  $\lambda'= -0.102$ ${\rm GeV}^{-1}$.

Some of the amplitudes, like $A_{2,1}^0$, 
$A_{2,4}^0$, $A_{2,5}^0$, all $A_{3,i}^0$  etc., contain the weak 
transition $D^*$ to $\pi$. 
This requires the $\pi$ to have very large momentum, which 
means that such graphs, which vanish in the heavy quark limit, 
give very small contributions.

Turning now to the presentation of the results we have to start with a discussion 
of the bremsstrahlung contribution (IB). 
In our model IB is given by diagrams 
$(IB)^o = \sum_{i}(A_{1,i}^0 + A_{2,i}^0 
+ C_{1,i}^0)$ for the $D ^0 \to K^- \pi^+ \gamma$ decay 
(the first two rows and the fourth row of Fig.3) 
and by diagrams $(IB)^+ =  \sum_{j}(A_{1,j}^+ + A_{2,j}^+ 
+ C_{1,j}^+ +C_{2,j}^+)$, i.e. the first two rows and rows 
5 and 6 of Fig. 1 for the $D^+ \to \bar K^0 \pi^+ \gamma$ decay. 
Now, in the limit of vanishing photon energy, 
the first two terms in the expansion of the IB amplitude in terms of the photon
energy, obey the Low theorem \cite{Low}. 
Although this is fulfilled theoretically, the question 
arises wheather the 
$D \to K\pi$ amplitude, as derived from our model,  
describes correctly the 
observed $D^+ \to {\bar K}^0 \pi^+$, 
$D^0 \to K^- \pi^+$ decays. We calculated the amplitudes of 
these decays 
using our model and we find that the branching ratios obtained 
with factorization approximation are 
$4.1\%$ and $17\%$ respectively, compared with observed 
branching ratios 
of $2.9\%$ and $3.9\%$ \cite{PDG}. 
It appears that although the model is reasonable for 
$D^+ \to \bar K^0 \pi^+$ (the $\Delta I = 3/2$ amplitude), it misses the amplitude of 
$D^0 \to K^- \pi^+$ by a factor of 2. 
On the one hand, this gives us a certain reassurance on the suitabilty 
of the model we use for calculating the radiative amplitudes. 
On the other hand, we shall 
perform also an alternative calculation, whereby the bremsstrahlung amplitudes of the model
are deleted from the total radiative amplitude and replaced by the 
"experimental amplitude". This procedure is undertaken in order to enforce 
the fulfilment of the Low theorem for our radiative amplitudes. 
Thus, we assume constant $D \to K \pi$ 
amplitude of correct magnitude to 
reproduce the observed rates of $D^+ \to \bar K^0 \pi^+$  
and $D^0 \to K^- \pi^+$, from which we calculate the 
bremsstrahlung (IB) amplitudes. 
These have the form of the first term in (\ref{amp1}) with 
constant $F_0$.  To this we add the $F_2$ terms of the 
magnetic transition, which is not affected by this procedure, 
as well as the parity-violating $F_1$ terms not belonging to
$(IB)^0$ and $(IB)^+$ diagrams. These $F_1$ terms then represent 
the direct electric transition of the radiative amplitude. We 
present results for both these alternative 
procedures. Although, the procedure based on the experimental 
$D \to K \pi$ amplitudes is apparently more reliable, we consider  
the "model" calculation to be of intrinsic value, setting out the 
ground for future calculations. 

There is one more item to be explained. We are interested in 
the role played by the vector mesons in these decays; obviously 
not in the direct $K \pi$ channel, which belongs to 
$D^0 \to {\bar K}^{*0} \gamma$ and was treated separately 
\cite{FPS2,L,BGHP}, rather as they appear as intermediate particles in VDM 
(e.g. diagrams $A_{3,4}^+$,  $C_{3,4}^+$,  $D_{2,3}^+$, 
$A_{3,3}^+$ and others) or in the crossed channels (e.g. 
diagrams $A_{4,1}^+$,  $D_{3,1}^+$, $D_{3,2}^+$ $C_{1,3}^0$,
$B_{3,2}^+$, and others). This is the main reason for our using an 
effective Lagrangian which contains the light vector mesons 
\cite{castwo}.
However, we calculate the radiative transitions also  
without including vector mesons in the Lagrangian, i.e. we drop all 
diagrams containing a double line in Figs 1-4  
(gauge invariance is 
maintained), which allows to elucidate their role 
in these decays. 

For the parity conserving part of the decays, 
representing the magnetic transition, we obtain 
\begin{equation}
BR(D^+ \to \bar K^0 \pi^+ \gamma)_{PC} = 2.0 \times 10^{-5}, 
\label{rpc+}
\end{equation}
\begin{equation}
BR(D^0 \to K^- \pi^+ \gamma)_{PC} = 1.4 \times 10^{-4}.
\label{rpc0}
\end{equation}
If we disregard the contribution of vector mesons, the rates are reduced to 
$BR(D^+ \to \bar K^0 \pi^+ \gamma)_{PC}^{no VM} = $ 
$3.0 \times 10^{-6}$ and 
$BR(D^0 \to K^- \pi^+ \gamma)_{PC}^{no VM} =  $ 
$ 6.6 \times 10^{-7}$. The decrease is sharper for the $D^0$ decay, 
since in this case the light vector mesons gave the dominant 
contribution to the rate, this is not 
the case for $D^+$ where such a contribution is doubly Cabibbo suppressed. 
The differential distribution for these transitions, as a function of 
$m_{12}^2 = (P- k)^2$, is given as the dashed line 
distribution 
in Fig. 5a, 5b, for these two decays. The distribution is mainly 
symmetrical, with the peak occuring at 
$k \simeq  400$ ${\rm MeV}$. Thus, this 
is the region in which the effect of the direct transition has best 
chance for detection. 

Turning to the parity - violating transitions we 
start with the procedure whereby we enforce the Low theorem by using Eq. (25).
Here we face the question of unknown phase between $F_0$ and $F_1$. 
We give therfore the results in terms of a range, limited by minimal 
and maximal interference between $F_0$ and $F_1$. 

Thus, we get for the branching ratios of the electric transitions,  
with $|F_0|$ determined experimentally, 
\begin{equation}
BR(D^+ \to \bar K^0 \pi^+ \gamma)_{PV,ex}^{k>50 MeV} = 
(3.6-3.8)\times 10^{-4}, 
\label{rpv+1}
\end{equation}
\begin{equation}
BR(D^+ \to \bar K^0 \pi^+ \gamma)_{PV,ex}^{k>100 MeV} = 
(2.3-2.5)\times 10^{-4}. 
\label{rpv+2}
\end{equation}
For the $D^0$ radiative decay we get 
\begin{equation}
BR(D^0 \to K^- \pi^+ \gamma)_{PV,ex}^{k>50 MeV} = 
(5.0-15)\times 10^{-4},
\label{rpv01}
\end{equation}
\begin{equation}
BR(D^0 \to K^- \pi^+ \gamma)_{PV,ex}^{k>100 MeV} = 
(2.6-11)\times 10^{-4}.
\label{rpv02}
\end{equation} 
The uncertainty in the $F_0/F_1$ phase is less of a problem 
in $D^+ \to \bar K^0 \pi^+ \gamma$ than in 
$D^0 \to K^- \pi^+ \gamma$. If we take the bremsstrahlung 
amplitude alone as determined 
from the knowledge  of $|F_0|$, disregarding the direct electric 
$F_1$ term, the above 
numbers are replaced by $3.6\times 10^{-4}$ and 
$2.3\times 10^{-4}$ for $D^+$ decay and $8.6 \times 10^{-4}$ 
and $5.5\times 10^{-4}$ for the $D^0$ decay. In Fig. 6 we also 
show the dependence of the branching ratio of the bremsstrahlung 
amplitude on the lower energy bound, for both 
decays. 
The contribution of the direct parity violating term (putting 
  $F_0 = 0$ ), is $BR(D^+ \to \bar K^0 \pi^+ \gamma)_{dir,PV } $ 
  $ =1.0 \times 10^{-5}$ and $BR(D^0 \to K^- \pi^+ \gamma)_{dir,PV} $ $
= 1.64 \times 10^{-4}$.

We checked also for the PV-transition the effect of the 
vector mesons. Using the Lagrangian of Refs. \cite{YAN,castwo} we found 
that in the PV-transition the effect of vector mesons is rather 
negligible; there is practically no change in (\ref{rpv+1}), 
(\ref{rpv+2}) and only a a narrowing of the range in 
(\ref{rpv01}) and (\ref{rpv02}), to bring it essentially to the values 
of pure bremsstrahlung we indicated after Eq. (\ref{rpv02}). 

We have calculated the decay rates also by using our model, for 
the whole radiative amplitudes 
i.e. using all graphs of Figs. 1-4. 
Comparing these results with those of Eq. (\ref{rpv+1}) - 
(\ref{rpv02}) gives an indication of the possible uncertainty of 
our model. 
We obtain 
\begin{equation}
BR(D^+ \to \bar K^0 \pi^+ \gamma)_{PV,model}^{k>50 MeV} = 
3.0\times 10^{-4}, 
\label{rpv+1m}
\end{equation}
\begin{equation}
BR(D^+ \to \bar K^0 \pi^+ \gamma)_{PV,model}^{k>100 MeV} = 
2.5\times 10^{-4}. 
\label{rpv+2m}
\end{equation}
For the $D^0$ radiative decay we get 
\begin{equation}
BR(D^0 \to K^- \pi^+ \gamma)_{PV,model}^{k>50 MeV} = 
2.3\times 10^{-3},
\label{rpv01m}
\end{equation}
\begin{equation}
BR(D^0 \to K^- \pi^+ \gamma)_{PV,model}^{k>100 MeV} = 
1.5\times 10^{-3}.
\label{rpv02m}
\end{equation} 
In Fig. 5c, 5d, we compare the rate of 
the decays $d\Gamma = d\Gamma_{PC} + d\Gamma_{PV}$ for the two 
alternative calculations concerning the PV part. In Fig. 5e, 5f we 
compare 
the rate of decay, $d\Gamma = d\Gamma_{PC} + d\Gamma_{PV}$ 
calculated from (\ref{amp1}) 
to the bremsstrahlung rate, to emphasize the 
feasibilty 
of detecting the direct emission. Finally, in Fig. 7 we 
present Dalitz plots for these decays.\\

 {\bf IV. DISCUSSION AND SUMMARY}\\

The calculation we presented is the first attempt to formulate a 
theoretical framework for decays of type $D\to K \pi \gamma$, with 
nonresonant $K\pi$. The calculational framework is the strong Lagrangian (12)-(15) used in the 
tree approximation,
and factorization for the weak matrix elements.
In the present article we treat the Cabibbo 
allowed decays and among these, only channels which have both inner 
bremsstrahlung and direct radiation components,  
$D^+  \to \bar K^0 \pi^+ \gamma$ and $D^0 \to K^- \pi^+ \gamma$, 
i.e. those which are the most likely ones for early detection. 
There is a third channel in this class, 
$D^0 \to \bar K^0 \pi^0 \gamma$, which has only a direct component in 
the radiative decay and will be discussed separately. 

Our results show that the relative expected strengths of the direct and 
bremsstrahlung components are of a magnitude which would permit 
the experimental determination of both, in next generation of 
experiments. This is important, since the direct amplitude provides 
information on the decay mechanism. In the radiative decay of 
$D^+$, the magnetic direct component amounts to about $6\%$ of the total rate 
(see Eqs. (33), (35)) and together with direct 
electric component which is of comparable magnitude (see Fig. 5a),  
dominate the decay spectrum in the region of high photon energies, say
above $k \simeq 250$ ${\rm MeV}$ (see Fig. 5f).
A similar situation occurs in the $D^0 \to K^- \pi^+ \gamma$  
transition, where the direct radiative decay containing both  
electric and magnetic parts which are 
of nearly equal
magnitude, amounts to over $30\%$ of the 
total radiative decay rate (see Eqs. (34) and (37)). 
The  numbers we mentioned are for $k > 50 $ ${\rm MeV}$, but as we 
stressed the region of high photon momenta beyond $200$  
${\rm MeV}$ is where the direct transition even dominates. 

The above large relative rates of direct/IB are somewhat similar to the 
occurences in $K_2^0 \to \pi^+ \pi^- \gamma$ and 
$K^+ \to  \pi^+ \pi^0 \gamma$, although in the D decays there is no 
apparant suppression of the original decays. 
This is most probably related to the mechanism of the decay provided by the heavy 
quark Lagrangian and the coupling constants involved, e.g. the 
rather large value of $|g|=0.59$ as recently determined.

We have checked the sensitivity of our results to various 
parameters we used. The uncertainty in the strong coupling 
$g=0.59 \pm 0.07$, 
may change our results for direct branching ratios by at most 
$15\%$. On the other hand, the uncertainty in $\lambda$ and 
$\lambda'$, and changing of sign of $\alpha_{1,2}$ is comparably
negligible. As to the values of $f_D$, if we vary it by a reasonable 
amount we can induce changes in the direct amplitudes by a few 
tens of 
percent. Althogeter, we feel that with the assumed calculational 
framework, the calculated direct amplitudes do not have 
uncertainties of more than $50\%$. 
Another feature which we neglect is the role of possible axial and 
scalar resonances, where the experimental situation is rather 
unsettled.

 As we explained in the text, the results (35) -(38) are obtained by 
 using the experimental $D \to K \pi$ amplitudes to calculate the inner 
 bremsstrahlung. If we  use the model 
 for doing it, we get the result exhibited in (39)- (42), which 
 do not differ from (35), (36), i.e. the 
 $D^+ \to \bar K^0 \pi^+ \gamma$ decay, but are larger by a factor of 
 about 2 in the amplitude in the case of  
 $D^0 \to K^- \pi^+\gamma$ decay. This is apparently related to the known difficulty
 of calculating the $D^0 \to K^- \pi^+$ amplitude in the factorization
 approximation; therefore, we consider the results given in 
 (37), (38) to be on a safer ground. 
 
 If we disregard the contribution of vector mesons to the direct part 
 of the radiative decays, the parity-conserving part of the amplitude 
 is considerably decreased, by one order of magnitude in the rate in 
  $D^+ \to \bar K^0 \pi^+ \gamma$ decay and by two orders of 
  magnitude in $D^0 \to K^- \pi^+ \gamma$. 
  On the other hand, their contribution is not felt in a significant way in 
in the parity - violating part of the amplitudes. 
In any case, the detection of the direct part of these decays at 
the predicted rates, will constitute a proof of the 
important role of the light vector mesons. The contribution of the 
vector mesons in the crossed channels is evident in the Dalitz plots of Fig.7.
We point out that the contribution of the vector mesons in $F_1$, as 
 evidenced 
from the relevant graphs, is wholy determined by the 
$\lambda$, $\lambda' $ 
couplings.

Figs. 5a, 5b give the expected spectra for the direct component, 
which would be detectable in the region of high photon energioes. 
For the $D^+$ decay, the direct electric and magnetic transitions 
are of comparable strength. This 
prediction of the model should be testable, as it shiffts the peak of 
the spectrum to $E_\gamma= 480$ ${\rm MeV}$, while if the magnetic transition
is dominant it should peak $E_\gamma= 400$ ${\rm MeV}$. In the 
$D^0 \to K^- \pi^+ \gamma$ decay  the magnetic and  
electric components are likewise of nearly equal  
size, again testable in the spectrum. It is worthwhile to point out that the relative values of the 
parity - conserving amplitudes is rather large 
$|A(D^0 \to K^- \pi^+ \gamma)|^2/|A(D^+ \to \bar K^0 \pi^+ \gamma)|^2$
$ \simeq 7$. The main reason for it are certain contributions 
(like $D_{1,1}^0$), which appear in $D^0$ decay but are doubly Cabibbo
forbidded in the $D^+$ decay. Also, as we pointed out, the 
$\Delta I = 1/2$ amplitude is larger than the $\Delta I = 3/2$ one
in the $D \to K \pi$ channels.

Finally, we wish to emphasize a most interesting implication of our calculation. 
When one compares the results obtained here for the 
radiative decays to nonresonant $K \pi$, with those previously obtained for the 
$D \to K^{*} \gamma$ [4,6,7,16],  
it emerges that the nonresonant channel is the more frequent one. 
Thus, while one expects for $D^+ \to K^{*+} \gamma$ a 
branching ratio (BR) of about 
$10^{-6}$ (this decay being doubly Cabibbo 
suppressed), the direct decay 
$D^+ \to \bar K^0 \pi^+ \gamma$ 
is expected to have a BR of $\simeq 3 \times 10^{-5}$ in our model. 
To this one should add the IB component, which brings 
its BR to about $4 \times 10^{-4}$ for $k > 50$   ${\rm MeV}$. 
The radiative decay $D^0 \to \bar K^{*0} \gamma$ is expected with 
BR of $0.5 \times 10^{-4}$. The nonresonant direct 
$D^0 \to K^- \pi^+ \gamma$ decay we investigated here, has a BR of 
$\simeq 3  \times 10^{-4}$ and including the IB component will occur 
with a rate $8  \times 10^{-4}$ for $k > 50$   ${\rm MeV}$.
The experimental verification of this systematics will provide a check 
for the suitabilty 
of the theoretical methods employed. 
  We should remark,however, at this point that we did not address the
possibility of the
decays $D\to R  + \gamma$ , where R is a higher $K \pi$ or 
$K \pi \pi$  resonance. To our
knowledge, there  
is no calculation available on this topic. Our expectation is that such
modes are at most
comparable in strength to the $D\to \bar K^* \gamma$ decay; 
preliminary data
from BELLE \cite{Belle}  indicate that this is the case in B decays.

There is another feature of interest 
which we mention. Since there is interference between the direct 
electric and 
the bremsstrahlung component of the amplitude, these decays can 
also be used to test  CP - violating effects in the amplitudes. 

We conclude by expressing the hope that the interesting features which 
these decays provide and were analyzed 
in this paper, will bring to an experimental 
search in the near future.

{\bf ACKNOWLEDGMENTS}\\

We thank our colleagues Y. Rosen, S. 
Tarem and P. Kri\v zan for stimulating 
discussions on experimental aspects of this investigation. 

The research of S.F. and A.P. was supported in part by the 
Ministry of Education, Science and Sport  
of 
the Republic of Slovenia. 
The research of P.S. was supported in part by Fund for 
Promotion of Research at the Technion. \\

\vspace{1cm}
{\bf APPENDIX A: THE DECAY AMPLITUDES FOR $D^+ \to \bar K^0 \pi^+ \gamma$ }\\

\noindent
Here we give the expressions for the sum of amplitudes in each row 
presented in Fig 1. The contributions which 
arise due to $O_1$ operator are:

$$A^{+}_{1}=-ie\frac{f_D f_\pi}{f_K}\frac{(v \cdot q+v \cdot k)p \cdot k}{v \cdot k}\;,$$
$$A^{+}_{2}=-ie\sqrt{\frac{M_s}{M}}\frac{f_{Ds} f_\pi}{f_K}g
\frac{p \cdot q - (v \cdot q) (v \cdot p) + v \cdot k (M-v \cdot p)}
{v \cdot p+\Delta}\frac{p \cdot k}{v \cdot k}\;,$$
$$A^{+}_{3}=ieg\sqrt{\frac{M_s}{M}}\frac{f_{Ds} f_\pi}{f_K}
\frac{(v \cdot k)(q \cdot k)(p \cdot k)}{v \cdot k+v \cdot p+\Delta}$$
$$\left(\frac{-1}{v \cdot k+\Delta} \left(2\lambda^{'}-\frac{\sqrt{2}}{2}\lambda g_v
\left(\frac{q_\omega}{3m_\omega^2}-
\frac{q_\rho}{m_\rho^2}\right)\right)+\frac{1}{v \cdot p+\Delta}\left(2\lambda^{'}-
\frac{\sqrt{2}}{3}\lambda g_v\frac{q_\phi}{m_\phi^2}\right)\right)\;,$$
$$A^{+}_{4}=i\sqrt{2} f_{Ds} f_\pi g_{\bar K^{0*}\bar K^0 \gamma} g_v \lambda e \sqrt{M_s M}
\frac{(v \cdot k)(q \cdot k)^2 }{(v \cdot (p+k)+\Delta)((p+k)^2-m_{K^*}^2+i\Gamma_{K^*}m_{K^*})}\;.$$

\vspace{1cm}
The contributions coming from $O_2$ operator 
are:

$$C^{+}_{1}=-ie\frac{f_D f_K}{f_\pi}\frac{(v \cdot p)(p \cdot k)}{v \cdot k}\;'$$
$$C^{+}_{2}=ie\frac{f_{D} f_K}{f_\pi}g
\frac{p \cdot q - (v \cdot p) (v \cdot q) + v \cdot k (M- v \cdot p)}
{v \cdot q+v \cdot k+\Delta}\frac{p \cdot k}{v \cdot k}\;,$$
$$C^{+}_{3}=-ieg\frac{f_{D} f_K}{f_\pi}
\frac{(v \cdot k)(q \cdot k)}{p \cdot k(v \cdot k+v \cdot q+\Delta)}$$
$$\left(\frac{-1}{v \cdot k+\Delta} \left(2\lambda^{'}-\frac{\sqrt{2}}{2}\lambda g_v\left(\frac{q_\omega}{3m_\omega^2}-
\frac{q_\rho}{m_\rho^2}\right)\right)-\frac{1}{v \cdot q+\Delta}\frac{\sqrt{2}}{2}\lambda g_v
\left(\frac{q_\omega}{3m_\omega^2}+\frac{q_\rho}{m_\rho^2}\right)\right)\;,$$
$$C^{+}_{4}=-i\sqrt{2}f_{D} f_K g_{\rho \pi \gamma} g_v\lambda e M
\frac{(v \cdot k)(p \cdot k)^2}{(v \cdot (q+k)+\Delta)((q+k)^2-m_{\rho}^2+i\Gamma_{\rho}m_{\rho})}\;.$$

\vspace{1cm}

\noindent
Next we give the expressions for the sum of amplitudes 
in each row 
presented in Fig 2. The contributions arising from the 
$O_1$ operator:

$$ B^{+}_{1}=2eM\frac{f_{D} f_\pi}{f_K}\lambda^,\left(\frac{1}{v \cdot k+\Delta}+g
\frac{f_{Ds}\sqrt{M_s}}{f_{D}\sqrt{M}}
\frac{v \cdot q}{v \cdot p+v \cdot k}\left(\frac{1}{v \cdot k+\Delta}+
\frac{1}{v \cdot p+\Delta}\right)\right)\;,$$
$$ B^{+}_{2}=-\sqrt{2M}e g_v g_{\bar K^{0*} \bar K^0 \gamma}f_\pi\frac{\alpha_1 M-\alpha_2 v \cdot q}
{(p+k)^2-m_{K*}^2-i\Gamma_{K^*}m_{K^*}}$$
$$+eg_{\rho \pi \gamma}g_\rho \frac{f_D}{f_K}\sqrt{M}\frac{\sqrt{M}+\frac{f_{Ds}}{f_D}
\sqrt{M_s}g \frac{M-v \cdot p}{v \cdot p+\Delta}}
{(k+q)^2-m\rho^2+i\Gamma_{\rho}m_{\rho}}\;,$$
$$ B^{+}_{3}=-\frac{1}{\sqrt{2}}e\frac{f_\pi}{f_K} \lambda g_v \left(\frac{q_\omega}{3m_\omega^2}-
\frac{q_\rho}{m_\rho^2}\right)\frac{1}{(v \cdot k+\Delta)}\left(M f_D +\sqrt{MM_s}f_{Ds}
\frac{v \cdot q}{v \cdot k+v \cdot p}\right)$$
$$ +\frac{\sqrt{2}}{3}\sqrt{MM_s}e\frac{f_{Ds} f_\pi}{f_K}g \lambda g_v \frac{q_\phi}{m_\phi^2}
\frac{v \cdot q}{(v \cdot p+\Delta)(v \cdot k+v \cdot p)}\;.$$

\vspace{1cm}

\noindent
The operator $O_2$ gives the following contributions:

$$D^{+}_{1}=-2eM\frac{f_{D} f_K}{f_\pi}\lambda^,\left(\frac{1}{v \cdot k+\Delta}+g
\frac{v \cdot p}{(v \cdot q+v \cdot k)(v \cdot k+\Delta)}\right)\;,$$
$$D^{+}_{2}=\sqrt{2M}eg_vg_{\rho \pi \gamma}f_K\frac{\alpha_1 M-\alpha_2 v \cdot p}
{(q+k)^2-m_{\rho}^2+i\Gamma_{\rho}m_{\rho}}$$
$$-eg_{\bar K^{0*} \bar K^0 \gamma}g_{K^*} 
\frac{f_D}{f_\pi}M\frac{1+g\frac{M-v \cdot q}{v \cdot q+\Delta}}
{(v \cdot q)((k+p)^2-m_{K^*}^2+i\Gamma_{K^*}m_{K^*})}\;,$$
$$D^{+}_{3}=\frac{1}{\sqrt{2}}Me\frac{f_D f_K}{f_\pi} \lambda g_v \left(\frac{q_\omega}{3m_\omega^2}-
\frac{q_\rho}{m_\rho^2}\right)\frac{1}{(v \cdot k+\Delta)}
\left(1+g\frac{v \cdot p}{v \cdot k+v \cdot q}\right)$$
$$ +\frac{1}{\sqrt{2}}Me\frac{f_{D} f_K}{f_\pi}g \lambda g_v 
\frac{v \cdot p}{(v \cdot q+\Delta)(v \cdot k+v \cdot q)}\left(\frac{q_\omega}{3m_\omega^2}+
\frac{q_\rho}{m_\rho^2}\right)\;.$$ 

\vspace{1cm}
{\bf APPENDIX B: THE DECAY AMPLITUDES FOR $D^0 \to K^- \pi^+ \gamma$ }\\

\noindent
The expressions for the sum of amplitudes (the $O_1$ operator) in 
each row exhibited in 
Fig 3. are:

$$ A^0_{1}=-iMe\frac{f_D f_\pi}{f_K}(v \cdot q+v \cdot k)\;, $$
$$ A^0_{2}=ie\sqrt{MM_s}\frac{f_{Ds} f_\pi}{f_K}g
\left(\frac{p \cdot k(v \cdot p-M)}{M(v \cdot p+\Delta)}
-\frac{p \cdot k(p \cdot q-(v \cdot p)(v \cdot q))}
{M(v \cdot p+\Delta)(v \cdot p+v \cdot k+\Delta)}\right.$$
$$ \qquad \left. +\frac{Mq \cdot k-Mp \cdot q+
M(v \cdot q)(v \cdot p)-(v \cdot q)(q \cdot k)}{M(v \cdot p+v \cdot k+\Delta)}\right)\;,$$
$$ A^0_{3}=ie\sqrt{M_s/M}\frac{f_{Ds} f_\pi}{f_K}g\frac{(v \cdot k)(p \cdot k)(q \cdot k)}{(v \cdot k+v \cdot p+\Delta)}
\left(\frac{2\lambda^{'}-\frac{\sqrt{2}}{3}\lambda g_v\frac{q_\phi}{m_\phi^2}}{(v \cdot p+\Delta)}
+\frac{\frac{\sqrt{2}}{2}\lambda g_v\left(\frac{q_\omega}{3m_\omega^2}+\frac{q_\rho}{m_\rho^2}\right)}
{v \cdot k+\Delta}\right)\;,$$
$$ A^0_{4}=i\sqrt{2}f_{Ds} f_\pi g_{K^{*} K \gamma} g_v \lambda e \sqrt{M_s M}
\frac{(v \cdot k)(q \cdot k)^2 }{(v \cdot (p+k)+\Delta)((p+k)^2-m_{K^*}^2+i\Gamma_{K^*}m_{K^*})}\;.$$

\vspace{1cm}
\noindent
The sum of amplitudes 
coming from operator $O_2$, shown in Fig. 3 as $ C^{0}_{1} $,  
is vanishing.

Next we present the expressions for the sums of amplitudes in each row 
shown in Fig. 4. The results for  the operator $O_1$ are:

$$  B^0_{1}=e\sqrt{M M_s}\frac{f_{Ds} f_\pi}{f_K}\frac{g(v \cdot q)}{(v \cdot p+\Delta)(v \cdot k+v \cdot p)}
\left(2\lambda^{'}+\frac{\sqrt{2}}{3}\lambda g_v\frac{q_\phi}{m_\phi^2} \right)$$
$$ +\frac{1}{\sqrt{2}}Me\frac{f_\pi}{f_K} \lambda g_v\frac{1}{(v \cdot k+\Delta)}\left(f_D 
\left(\frac{q_\omega}{3m_\omega^2}+\frac{q_\rho}{m_\rho^2}\right)+f_{Ds}\sqrt{\frac{M_s}{M}}
\left(\frac{q_\omega}{3m_\omega^2}-\frac{q_\rho}{m_\rho^2}\right)\frac{gv \cdot q}{(v \cdot k+v \cdot p)}\right)\;,$$
$$B^0_{2}=-\sqrt{2M}e g_v g_{\bar K^{*} \bar K^\gamma}f_\pi\frac{\alpha_1 M-\alpha_2 v \cdot q}
{(p+k)^2-m_{K*}^2+i\Gamma_{K^*}m_{K^*}}$$
$$ +eg_{\rho \pi \gamma}g_\rho \frac{f_D}{f_K}\frac{M+\frac{f_{Ds}}{f_D}
\sqrt{M_sM}g \frac{M-v \cdot p}{v \cdot p+\Delta}}{(k+q)^2-m_\rho^2+i\Gamma_{\rho}m_{\rho}}\;.$$

\vspace{1cm}

\noindent
Finally,  the sums of amplitudes in each row 
due to the operator $O_2$:

$$ D^0_{1}=\frac{M}{\sqrt{2}}e\lambda g_v f_D \left(\frac{q_\omega}{3m_\omega^2}+\frac{q_\rho}
{m_\rho^2}\right)\frac{1}{v \cdot k+\Delta}
\left(1+ \frac{m_{\bar K^0}^2}{(p+q)^2-m^2_{\bar K^0}}\right)\;,$$
$$ D^{0}_{2}=Me\frac{f_D}{f_\pi}g_{K}g_{K K^{*} \gamma}\frac{1}{(p+k)^2-m^2_{K^*}+im_{K^*}\Gamma_{K^*}}+
Me\frac{f_D}{f_K}g_{\rho}g_{\rho\pi\gamma}\frac{1}{(q+k)^2-m^2_{\rho}+im_{\rho}\Gamma_{\rho}}\;, $$
$$ D^0_{3}=-\frac{e}{2}f_D f_K \frac{m^2_{K^{*}}}{g_{K^{*}}}
\frac{M^3}{(M^2-m^2_{\bar K^0})}\left(\frac{g_{K^{*}K \gamma}}{(p+k)^2-m^2_{K^*}+im_{K^*}\Gamma_{K^*})}
+\frac{g_{\rho \pi\gamma}}{(q+k)^2-m^2_{\rho}+im_{\rho}\Gamma_{\rho})}\right)\;.$$

\newpage

\begin{figure}
\begin{center}
\includegraphics[width=19cm]{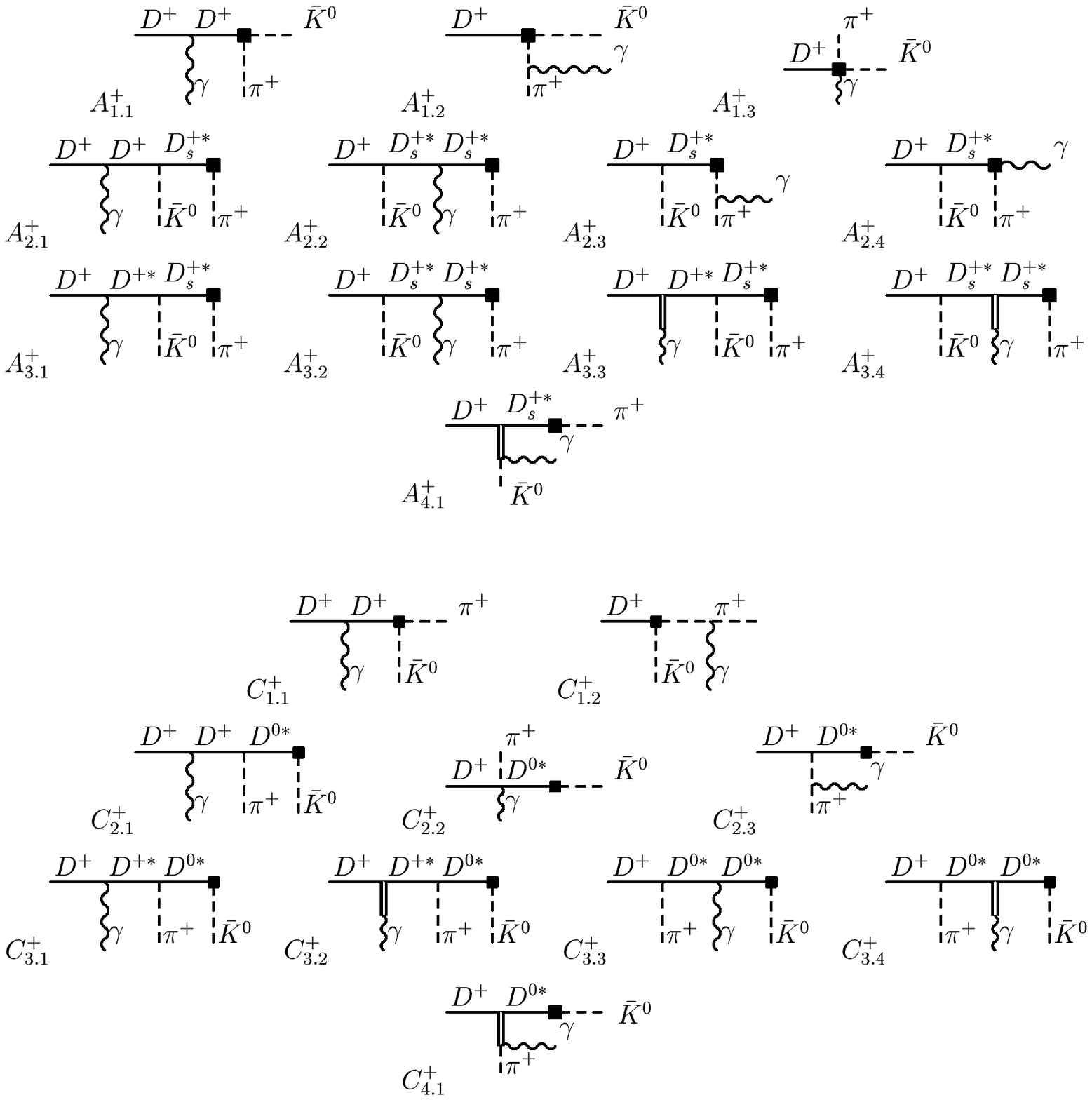}
\caption{Feynman diagrams contributing to the formfactor $F_1$ of the
$D^+ \to \bar K^0 \pi^+ \gamma$ decay. Diagrams denoted by $A^{+}_{i,j}$ 
($C^{+}_{i,j}$) come from the operator $O_1$ ($O_2$). Sum of the 
contributions of each row is gauge invariant. 
In diagrams $A^{+}_{3,1}$, $A^{+}_{3,2}$, $C^{+}_{3,1}$ and $C^{+}_{3,2}$ 
the photon couples to the heavy mesons with strength $\lambda^{'}$. }
\end{center}
\end{figure}

\begin{figure}
\begin{center}
\includegraphics[width=19cm]{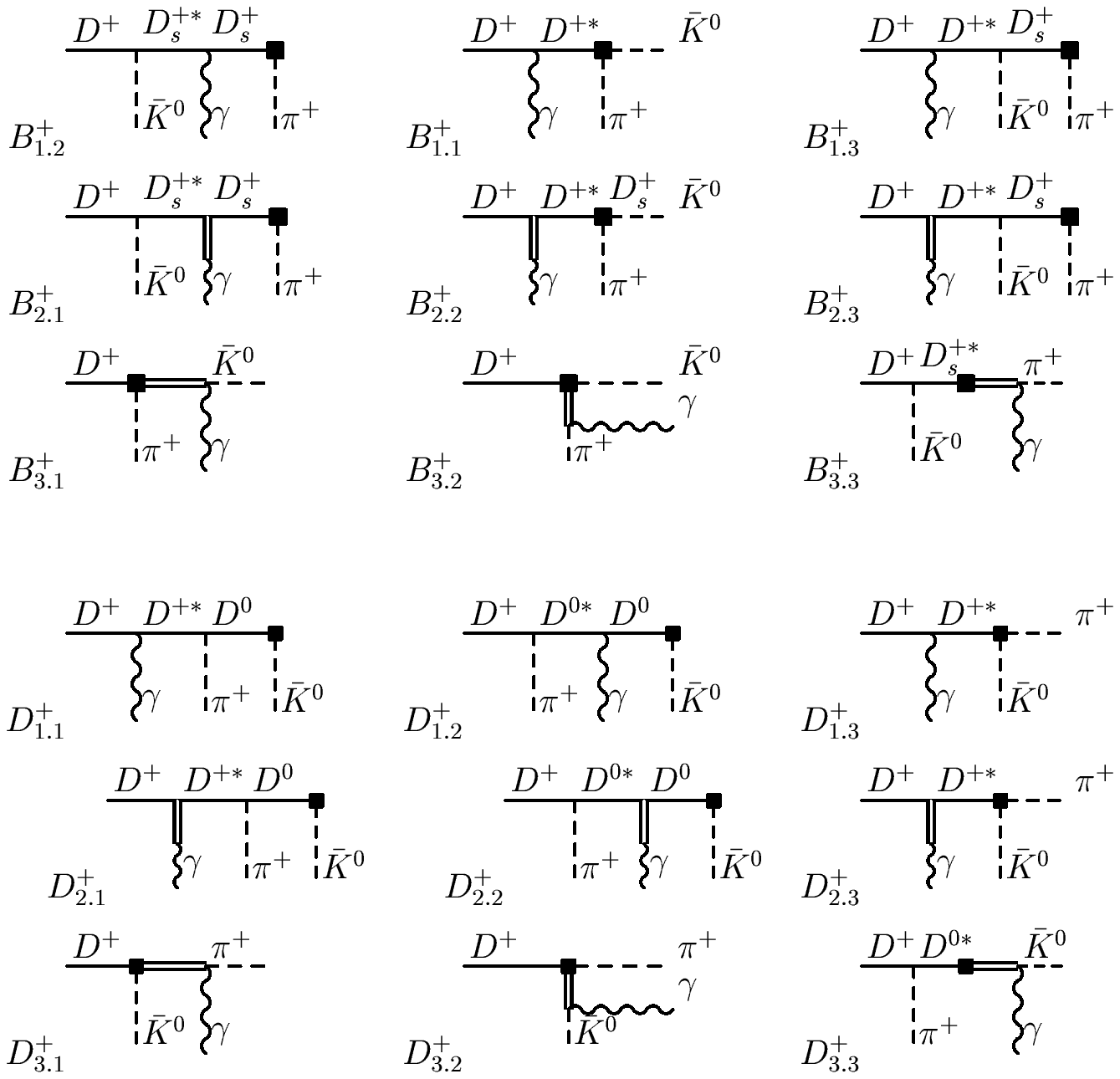}
\caption{Feynman diagrams contributing to the formfactor $F_2$ of the
$D^+ \to \bar K^0 \pi^+ \gamma$ decay. Diagrams denoted by $B^{0}_{i,j}$ 
($D^{0}_{i,j}$) come from the operator $O_1$ ($O_2$). }
\end{center}
\end{figure}

\begin{figure}
\begin{center}
\includegraphics[width=19cm]{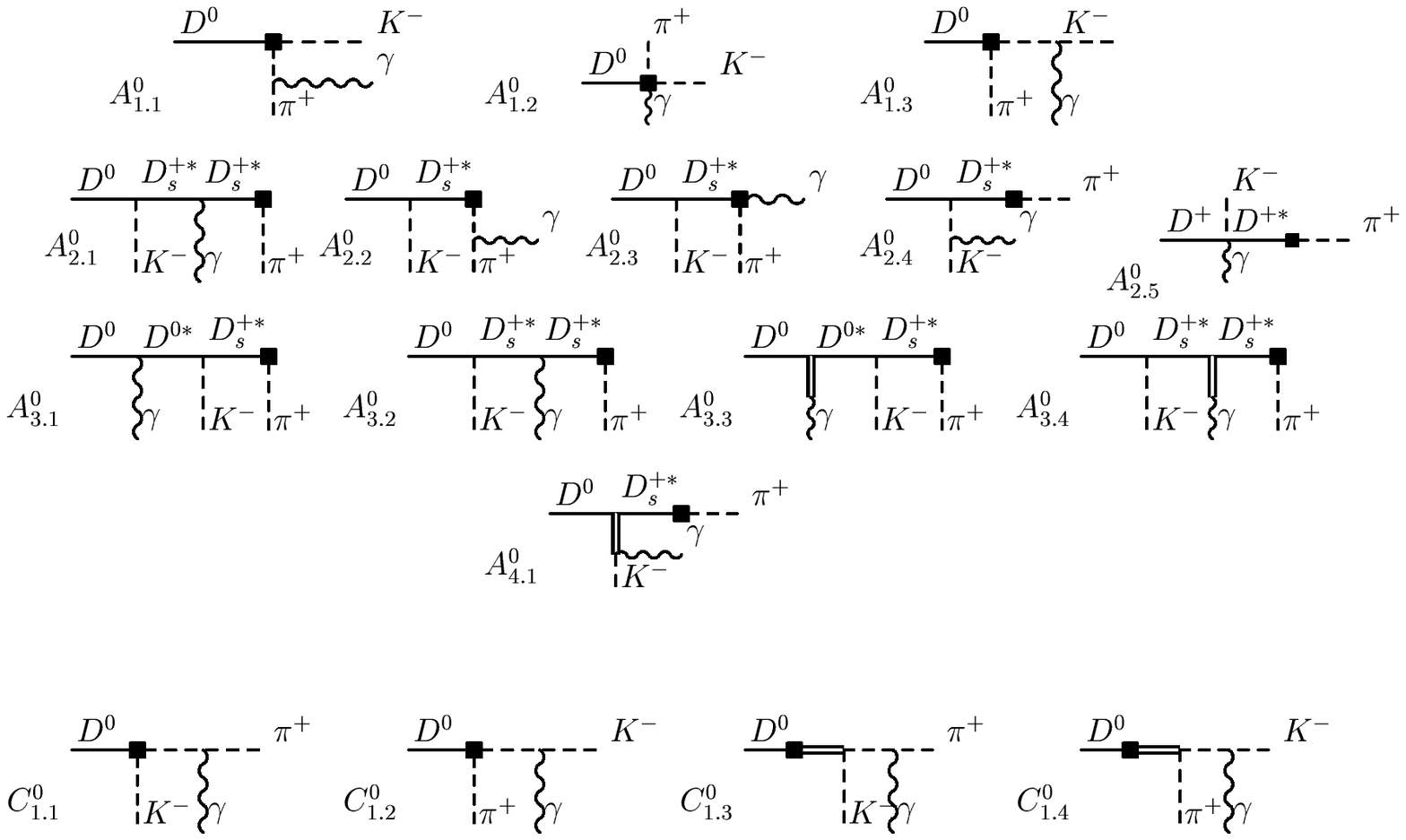}
\caption{Feynman diagrams contributing to the formfactor $F_1$ of the
$D^0 \to K^- \pi^+ \gamma$ decay. Diagrams denoted by $A^{+}_{i,j}$ 
($C^{+}_{i,j}$) come from the operator $O_1$ ($O_2$). Sum of the 
contributions of each row is gauge invariant.
In diagrams $A^{0}_{3,1}$ and $A^{0}_{3,2}$ the photon couples to the heavy 
mesons with strength $\lambda^{'}$. }
\end{center}
\end{figure}

\begin{figure}
\begin{center}
\includegraphics[width=19cm]{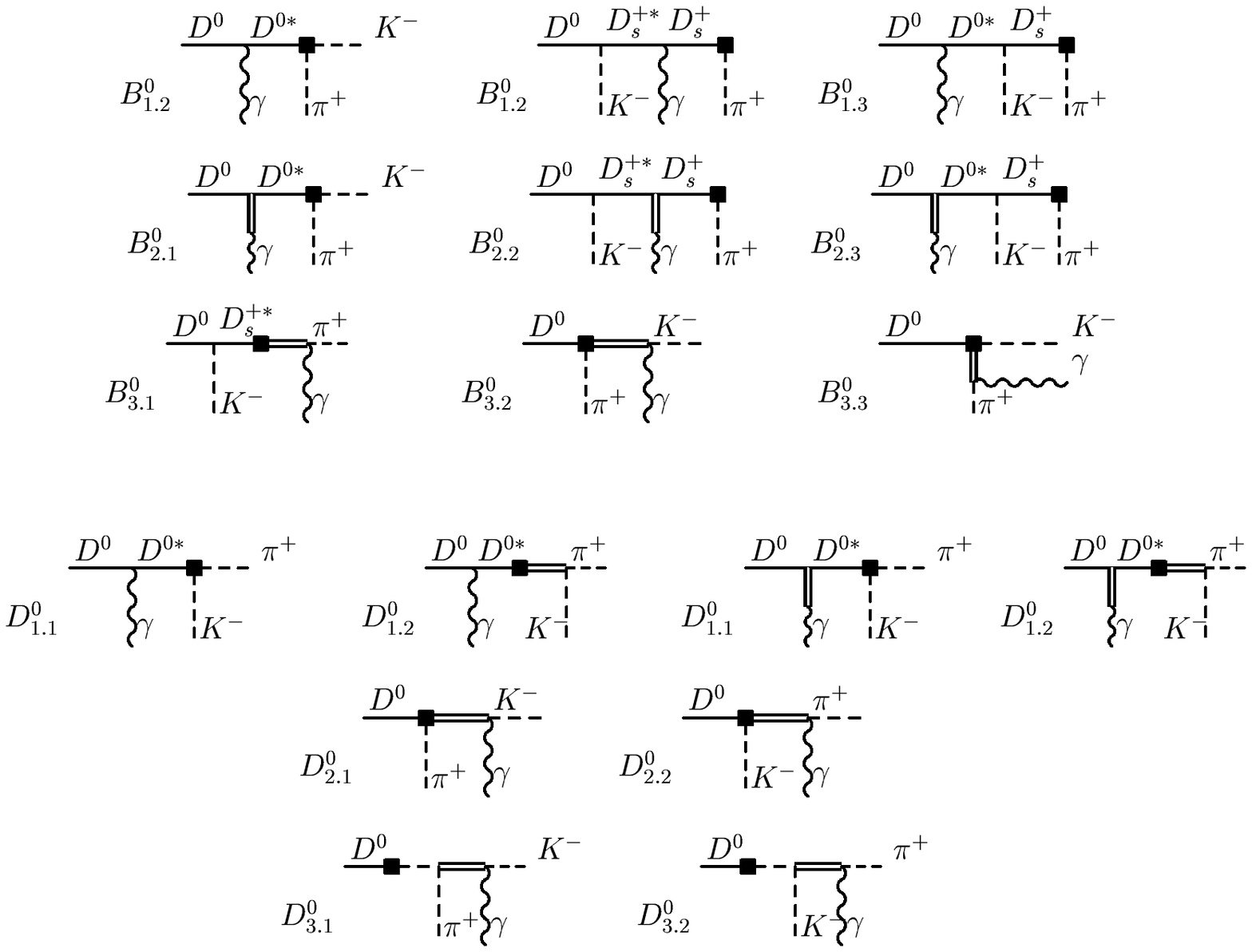}
\caption{Feynman diagrams contributing to the formfactor $F_2$ of the
$D^0 \to K^- \pi^+ \gamma$ decay. Diagrams denoted by $B^{+}_{i,j}$ 
($D^{+}_{i,j}$) come from the operator $O_1$ ($O_2$). }
\end{center}
\end{figure}

\begin{figure}
\includegraphics[width=8cm]{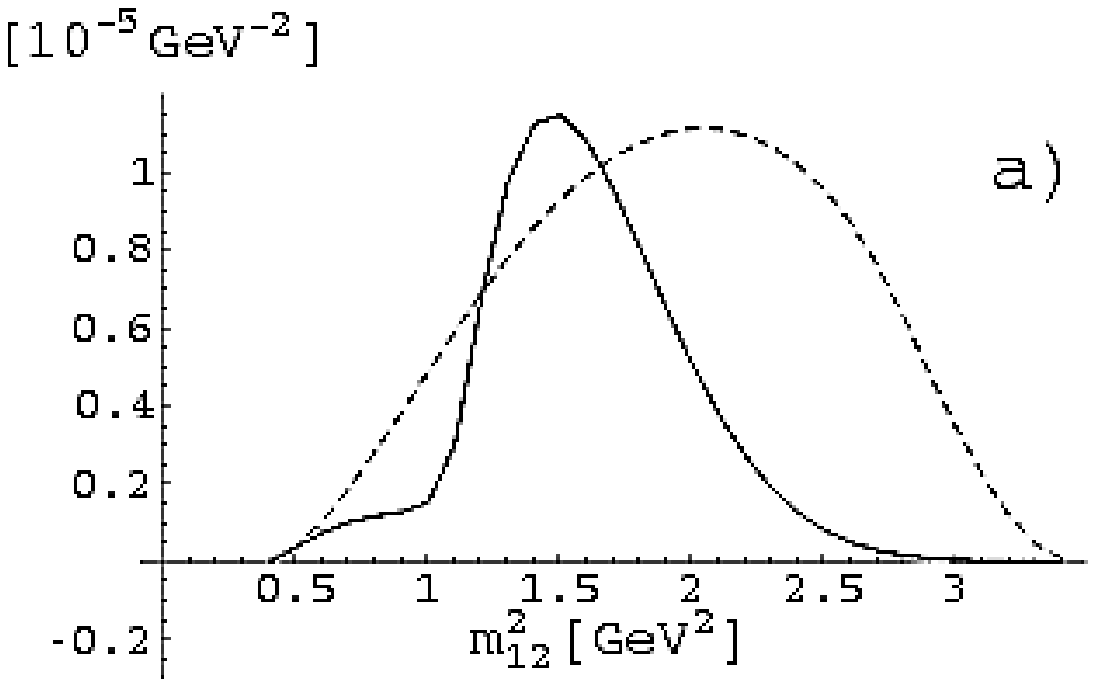}
\includegraphics[width=8cm]{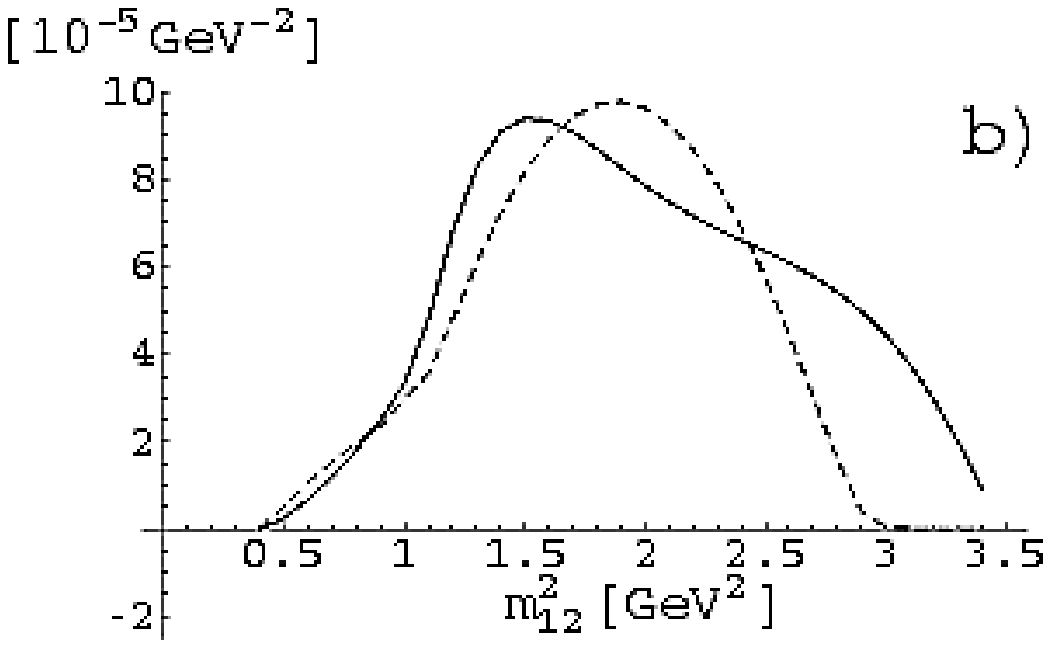}
\includegraphics[width=8cm]{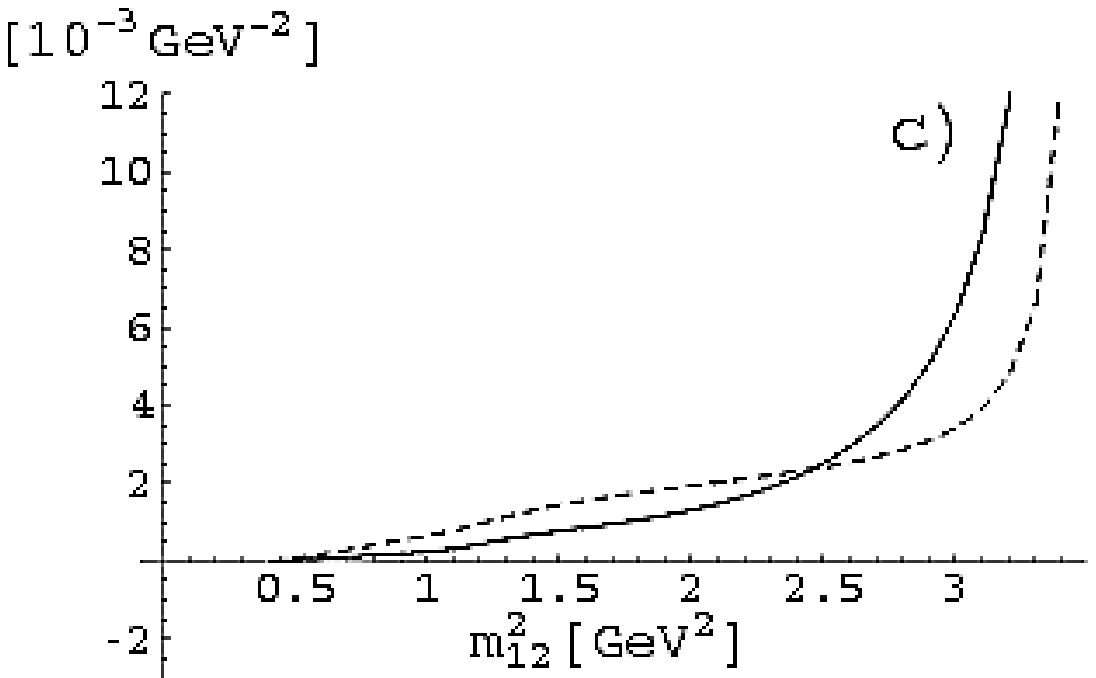}
\includegraphics[width=8cm]{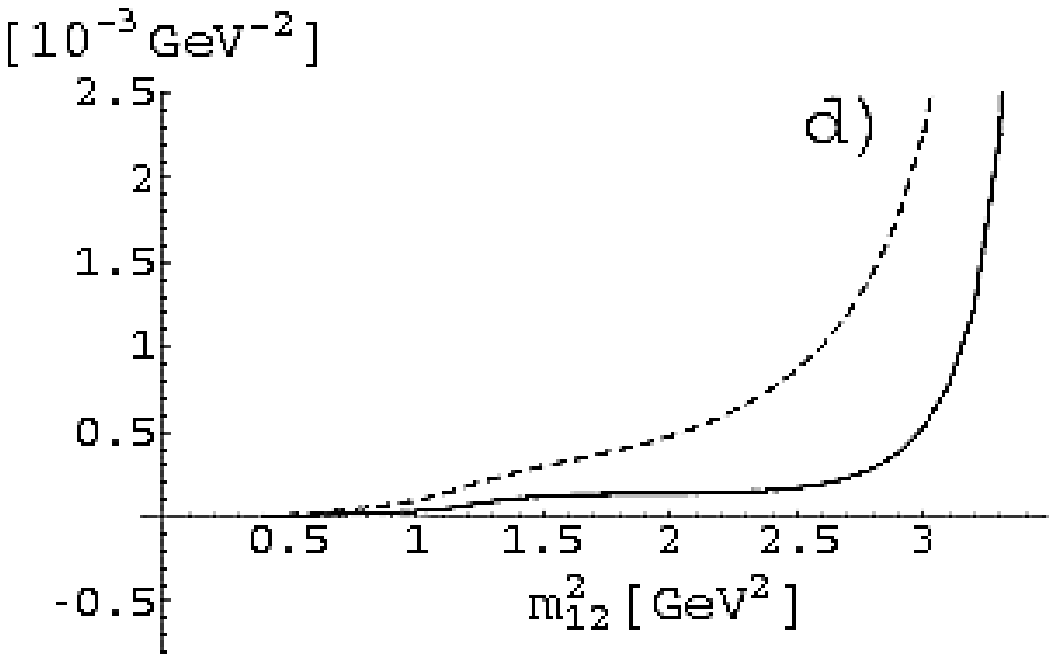}
\includegraphics[width=8cm]{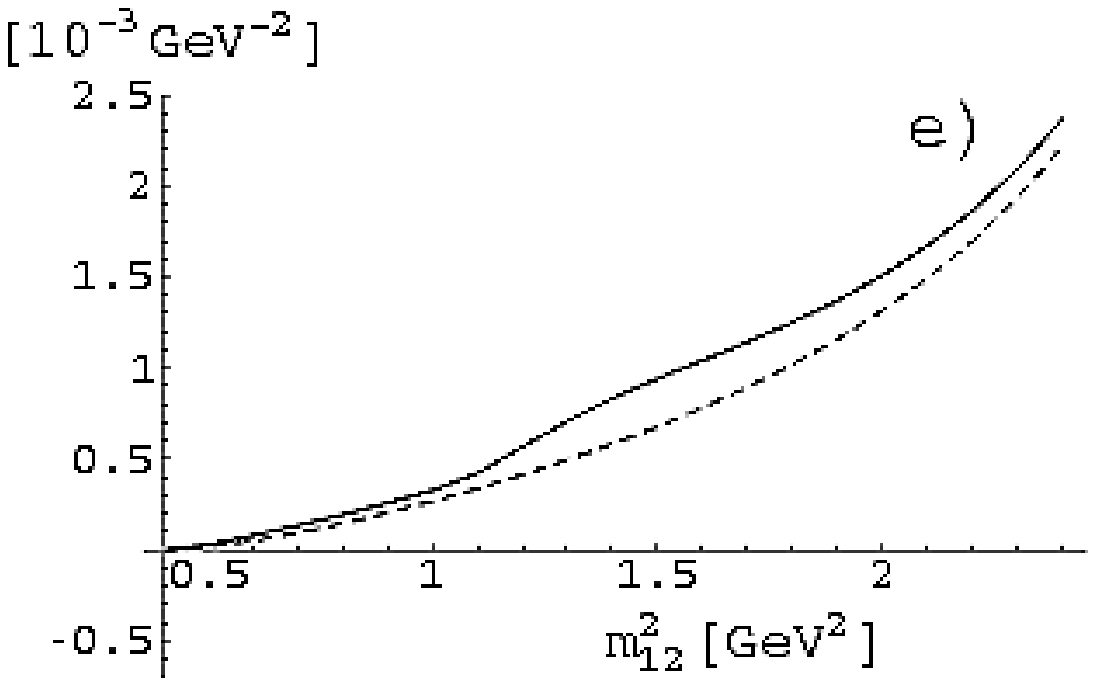}  
\includegraphics[width=8cm]{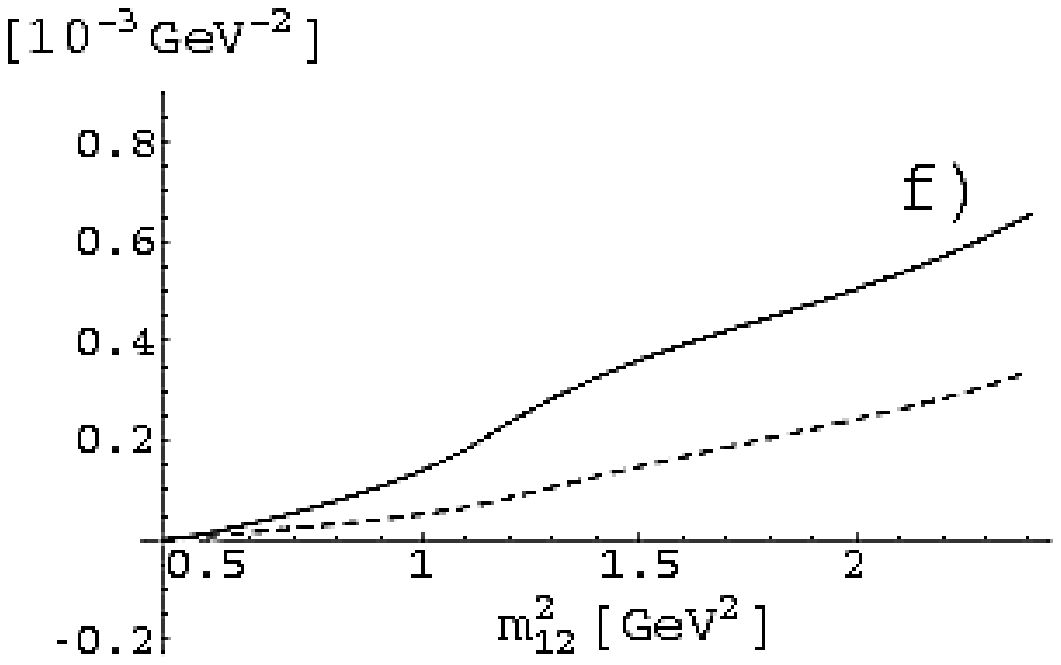}

\caption{$\frac{1}{\Gamma_{total}}\frac{d\Gamma}{d m_{12}}$
for the decay $D^+ \to \bar K^0 \pi^+ \gamma$
(left) and $D^0 \to K^- \pi^+ \gamma$ (right).
Above: Direct parity - conserving (dashed line) and parity - violating
(putting $F_0=0$) (full line) terms. Middle: 
$\frac{1}{\Gamma_{total}}\frac{d\Gamma}{d m_{12}}$ with $\Gamma$ 
containing the full decay amplitudes, for model (dashed line) and 
model+exp. (full line). For the latter, maximal $F_0$, $F_1$ interference 
is exhibited.
Below: $\frac{1}{\Gamma_{total}}\frac{d\Gamma}{d m_{12}}$ with $\Gamma$ 
for the radiative decay calculated from model+exp. (full - line) compared to pure bremsstrahlung emission (dashed - line).}
\end{figure}

\begin{figure}
\includegraphics[width=8.5cm]{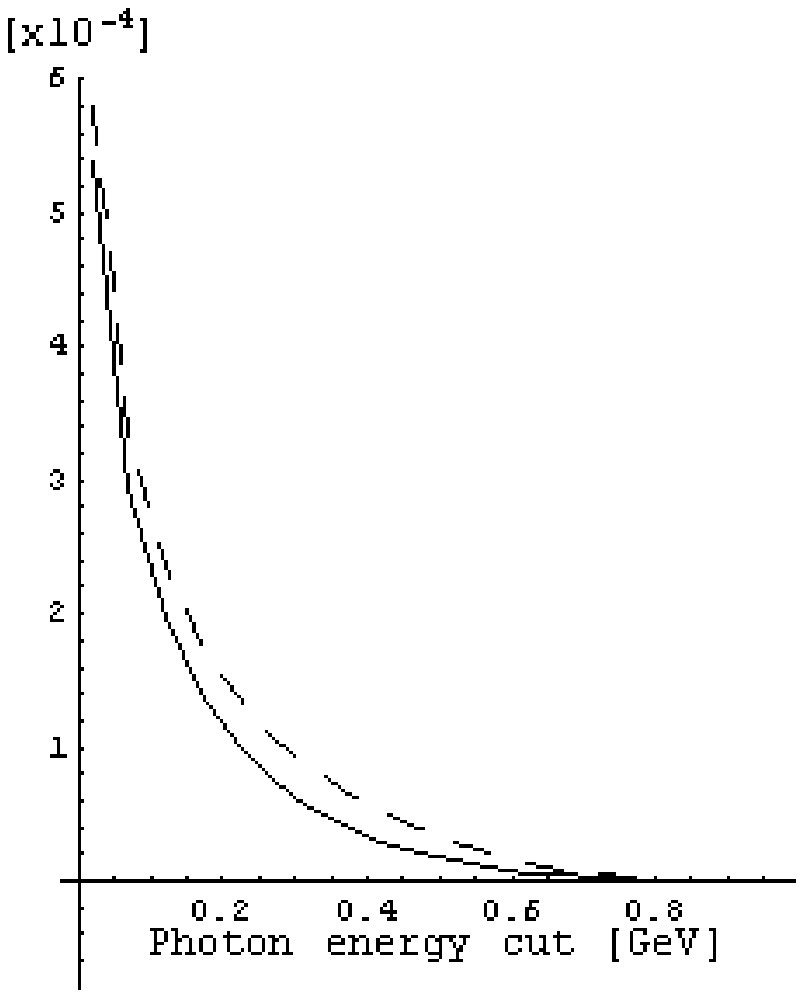}
\includegraphics[width=8.5cm]{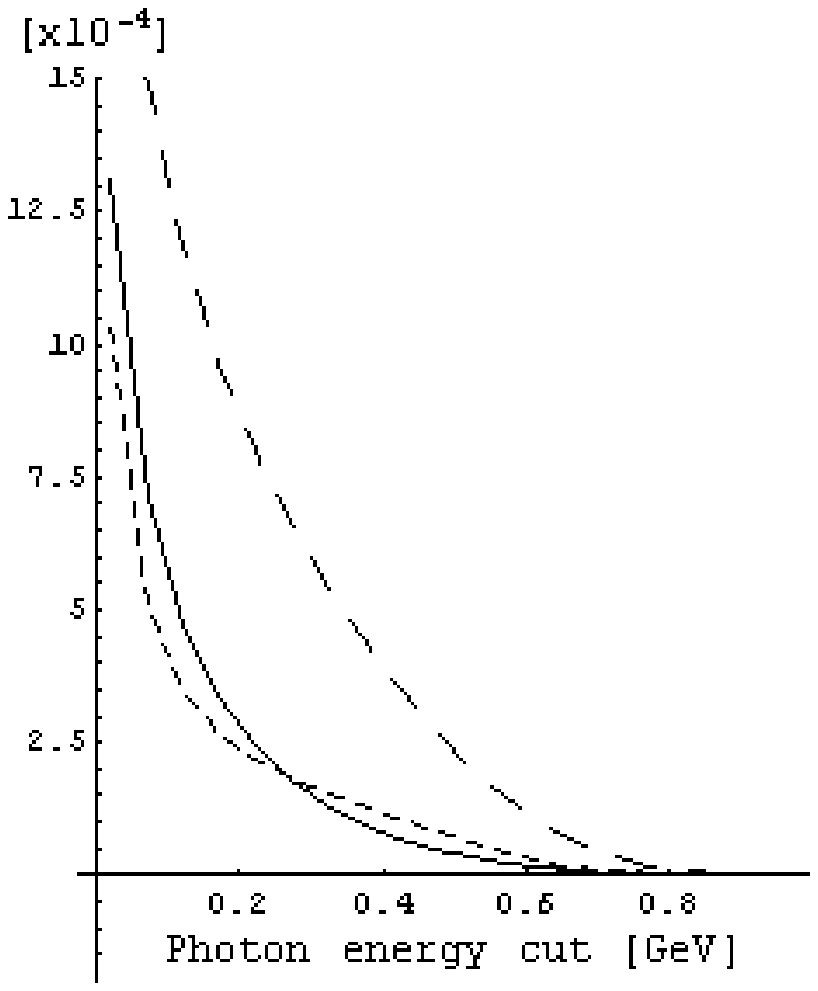}
\caption{Branching ratio of radiative decays. 
Left: Decay $D^+ \to \bar K^0 \pi^+ \gamma$
Right: Decay  $D^0 \to K^- \pi^+ \gamma$
Full line: Bremsstrahlung only.
Long dashed line: all contributions included and pozitive 
$F_0$, $F_1$ interference;
short dashed line: negative interference }
\end{figure}

\begin{figure}[h]
\begin{center}
\includegraphics[width=7cm]{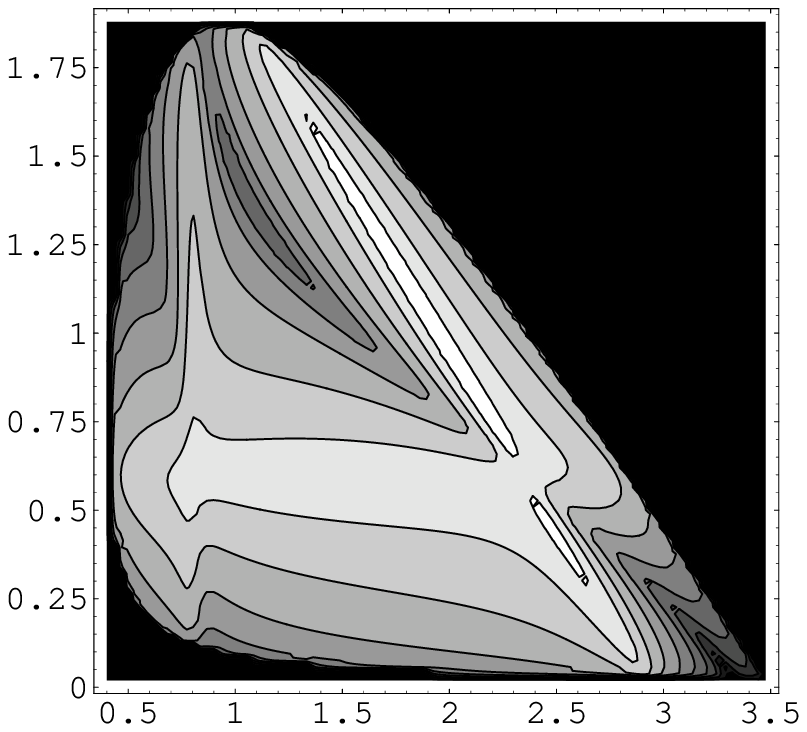}
\includegraphics[width=7cm]{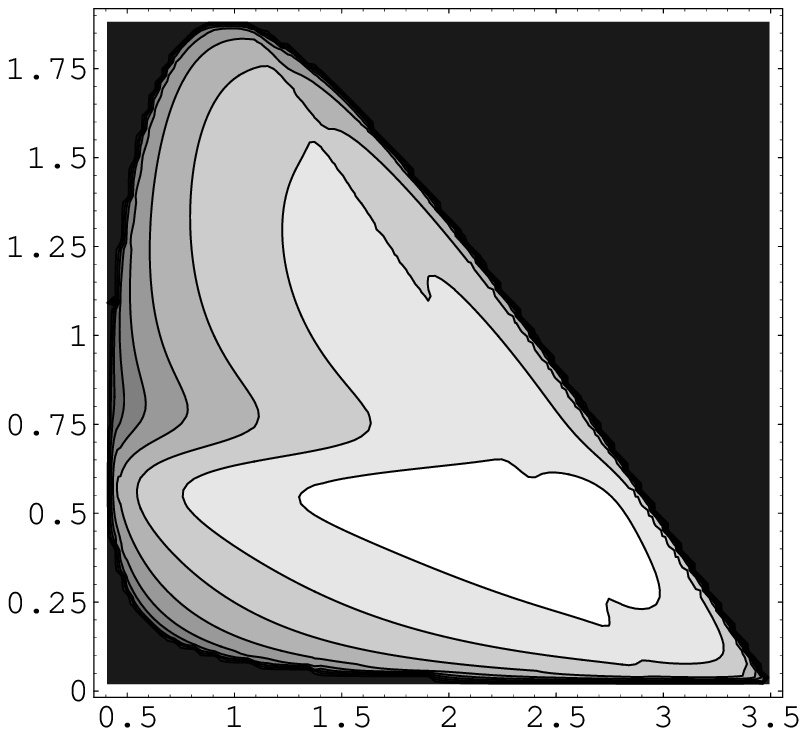}
\includegraphics[width=7cm]{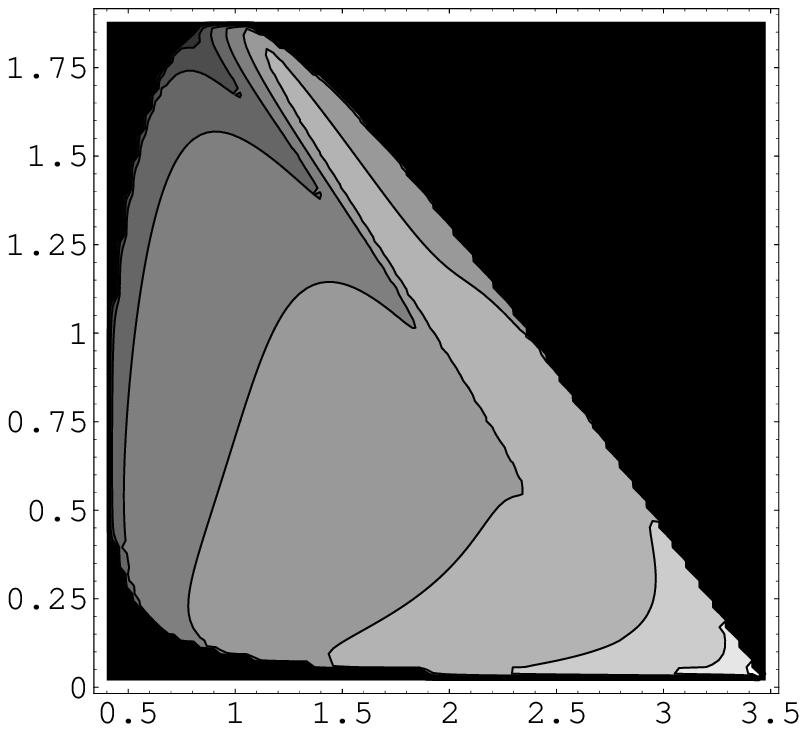}
\includegraphics[width=7cm]{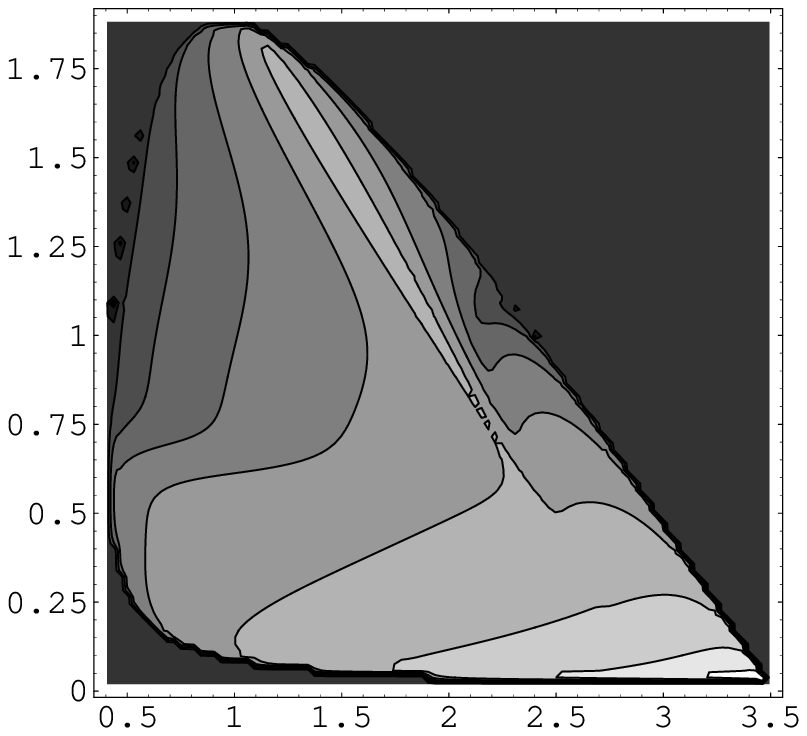}

\caption{Dalitz plot of the parity conserving (above) and parity violating
part (below) of the 
$D^0 \to K^- \pi^+ \gamma$ (left) 
and $D^+ \to \bar K^0 \pi^+ \gamma$ decay (right).
With gray levels on contour plot (left) and on z axis on the 3D plot (right) 
we present $\frac{2}{G_F} d\Gamma / (dm_{12}dm_{23})$ in the logarithmic scale.
Invariant mass $m_{12}=\sqrt{(P-k)^2}$
is plotted on the x axis and $m_{23}=\sqrt{(P-p)^2}$
on the y axis of contour plot. The x and y axes of 
the 3D plot are labeled. }
\end{center}
\end{figure}

\end{document}